\documentclass[12pt]{iopart}

\usepackage{graphicx}
\usepackage{amsopn}

\begin{document}

\title[The formation of a full shock]{The evolution of a slow electrostatic shock into a plasma shock mediated by electrostatic turbulence}

\author{M.E. Dieckmann$^1$, G. Sarri$^2$, D. Doria$^2$, H. Ahmed$^2$ and M. Borghesi$^2$}
\address{1. Department of Science and Technology (ITN), Link\"opings University,
 Campus Norrk\"oping, SE-60174 Norrk\"oping, Sweden}
\address{2. Centre for Plasma Physics (CPP), Queen's University Belfast, BT7 1NN, UK}
\ead{mark.e.dieckmann@liu.se}

\begin{abstract}
The collision of two plasma clouds at a speed that exceeds the ion acoustic speed can result in the formation of shocks. This phenomenon is observed not only in astrophysical scenarios such as the propagation of supernova remnant (SNR) blast shells into the interstellar medium, but also in laboratory-based laser-plasma experiments. These experiments and supporting simulations are thus seen as an attractive platform for the small-scale reproduction and study of astrophysical shocks in the laboratory. We model two plasma clouds, which consist of electrons and ions, with a 2D PIC simulation. The ion temperatures of both clouds differ by a factor of 10. Both clouds collide at a speed, which is realistic for laboratory studies and for SNR shocks in their late evolution phase like that of RCW86. A magnetic field, which is orthogonal to the simulation plane, has a strength that is comparable to that at SNR shocks. A forward shock forms between the overlap layer of both plasma clouds and the cloud with the cooler ions. A large-amplitude ion acoustic wave is observed between the overlap layer and the cloud with the hotter ions. It does not steepen into a reverse shock, because its speed is below the ion acoustic speed. A gradient of the magnetic field amplitude builds up close to the forward shock as it compresses the magnetic field. This gradient gives rise to an electron drift that is fast enough to trigger an instability. Electrostatic ion acoustic wave turbulence develops ahead of the shock. It widens its transition layer and thermalizes the ions, but the forward shock remains intact.    
\end{abstract}

\maketitle

\section{Introduction}

The collision between two plasma clouds may trigger the creation of shock waves, if the relative velocity between the two plasmas exceeds the ion-acoustic speed at the point of collision. This scenario is of particular relevance in astrophysics, since it occurs during the propagation of supernova remnants (SNR) in space. The dense and hot blast shell of a  SNR is in fact propagating through the interstellar medium (ISM), a much colder and more rarefied medium. The low-collisionality of the ISM (typical temperatures and densities of the order of the eV and of a particle per cm$^{-3}$, respectively) \cite{Ferriere01} guarantees that the dynamics of the shock waves is predominantly governed by electromagnetic fields; the shock is thus referred to as collisionless. This is not the only scenario in which collisionless shocks can be generated: other possible examples are represented by the bow-shock region (see, for instance, Ref. \cite{Bale}), and the atmosphere of microquasars \cite{Mirabel}. Due to the obvious difficulty in directly probing the microphysical conditions around the shock in an astrophysical scenario, dedicated effort has been recently devoted to the creation of comparable, smaller-scale reproductions in the laboratory. 

A particularly appealing scenario is offered by the interaction of an intense laser pulse with a solid target. The impact of the laser onto the solid heats up a significant population of electrons at the critical surface, which can reach temperatures of the order of a few MeV. The hotter electrons are able to indefinitely escape from the target, setting, by space charge separation, a net electrostatic field that starts to accelerate ions \cite{Gitomer86,Maksimchuk00,dHumieres05,Macchi13}. Due to the favourable charge to mass ratio, hydrocarbon ions resulting from surface impurities are the first to be accelerated followed, at a later time, by ions of the solid itself. These ions expand into the surrounding medium in the form of a rarefaction wave \cite{Crow75,Schamel87,Mora05,Quinn12}, which is characterized by a decreasing density and an increasing velocity as we move further from the source.  Meanwhile, x-rays emanating from the laser interaction point induce photo-ionisation of the low-density gas embedding the target. This induces a low-density, and low-temperature stationary ambient plasma through which the rarefaction wave is forced to propagate. 

Plasma shocks can form if the relative speed between the rarefaction wave and the ambient plasma exceeds the ion acoustic speed at the location where the densities of the rarefaction wave and the ambient medium are similar. If the plasma is unmagnetized or weakly magnetized and if the shock speed is below a few percent of the speed of light $c$, then electrostatic shocks and double layers form \cite{Hershkowitz81}. The shock speed depends on the details of the phase space distribution of the rarefaction wave and on how its density compares to that of the ambient medium. Simulations have demonstrated that a shock forms well behind the front of the rarefaction wave and that it expands away from the ablated target \cite{Silva04,Sarri11,SarriPRL}. The shock reflects a significant fraction of the ions of the ambient plasma, but some of them can also cross the shock boundary and move downstream. The accumulation of incoming upstream ions in the downstream region implies that the density behind the shock is locally increased compared to the density of the rarefaction wave and that the mean speed of the downstream plasma is reduced compared to the local speed of the rarefaction wave. The latter follows from momentum conservation. A reverse shock, which moves towards the target, is likely to form if the difference between the mean speed of the downstream region and that of the successive rarefaction wave, which corresponds to the laser-ablated plasma, exceeds the sound speed. This process has been observed experimentally \cite{Ahmed13}. 

It has been proposed in Ref. \cite{Remington99} to study the forward shocks, which can now be generated routinely in the laboratory \cite{Chen07,Romagnani08,Nilson09,Morita10,Kuramitsu11,Kugland12,Habersberger12,Fox13,Huntington14}, to better understand the properties of astrophysical shocks like the ones that form between the blast shell of a supernova remnant \cite{Woosley86} and the interstellar medium \cite{Ferriere01}. The rarefaction wave, which expands away from the laser-ablated target, would take the role of the supernova blast shell while the ambient medium would correspond to the ISM plasma. The possibility of studying astrophysical shocks in the form of a laboratory experiment is intriguing. However, experimental constraints exist that need to be addressed when comparing the results of laboratory experiments to astrophysical observations. 

Laser-driven shock waves are usually observed only for a short time after they have formed and transient effects arising from the initial conditions may still be important. One has to keep in mind that what is commonly referred to as an electrostatic shock \cite{Hershkowitz81} and tends to form quickly \cite{ForslundA,ForslundB,Dieckmann13a} is not necessarily what is called a shock in an astrophysical context. The latter implies a full thermalization of the downstream plasma. An electrostatic shock is characterized by an electric field that points along the shock normal; ions can not be deflected and heated perpendicularly to this field as they cross the shock and no full thermalization is possible. Ion thermalization can be accomplished by an electrostatic shock only through the ion acoustic instability that develops ahead of it \cite{Jackson60,ForslundC,Karimabadi91,Kato10,Dieckmann13b}. However, the back-reaction of the turbulence on the electrostatic shock may destroy it \cite{Kato10}.

Let us compare the laboratory- and astro-plasma parameters. The ambient medium for SNR shocks is the ISM. A significant fraction of it are neutral atoms, molecules or dust. SNR shocks thus plough through a medium that is either charge neutral (atomic material) or through a proton plasma with a temperature of the order eV. In the laboratory, the ambient plasma consists of fully ionized nitrogen and oxygen ions. Their characteristic temperature is of the order of hundreds of eV \cite{Ahmed13} and thus much higher than that ahead of SNR shocks. During the typical observational window of laser-driven shocks, the ion temperature downstream of the shock may not have reached a steady state and it is still determined by the temperature of the laser-generated blast shell. The electrons of the ambient plasma in the laboratory have temperatures of the order of a kilo-electron Volt (keV). This high temperature develops firstly because some of the laser-heated electrons can escape from the target and, secondly, because the target's secondary X-ray emission produces hot electrons as it ionizes the residual gas. The electrons of the warm ionized ISM have eV temperatures far from SNR shocks and those of the dilute hot ionized ISM have keV temperatures. The source of the latter are probably SNR shocks. A dense population of electrons with keV temperatures and a dilute population of cosmic ray electrons with higher energies exist close to SNR shocks \cite{Koyama95,Helder09,Raymond09}. Although the similar temperature of the bulk electrons is encouraging, we have to keep in mind that SNR shocks are faster than those we obtain in the laboratory unless the laser pulse is ultra-intense \cite{Nilson09}. The faster expansion speed implies that the Mach number of most SNR shocks with respect to the ion acoustic speed is larger than that of the shocks, which are generated in the laboratory. Usually, such fast shocks are at least partially mediated by self-generated magnetic fields \cite{Kato10,Stockem14}.

Many laboratory studies have addressed the slower electrostatic unmagnetized shocks, which have a narrow transition layer with a width of the order of an electron skin depth. Such structures can be detected at a high spatio-temporal resolution by means of the proton radiography technique \cite{Koehler68,Borghesi02,Sarri10}. It measures the deflection of probing protons by the electromagnetic field in the plasma. Its contrast is determined by the amplitude and scale of the field variations, which depend on the structure of the shock. The ISM into which SNR shocks expand is magnetized and laboratory experiments that address SNR shocks will aim at introducing an ambient magnetic field.

A perpendicular magnetic field in the shock transition layer can give rise to a relative motion between electrons and ions through gradient drifts, which could trigger the lower hybrid drift instability \cite{Brackbill84,Daughton03,Daughton04} or the electron cyclotron drift instability \cite{Forslund72,Umeda12,Belyaev05} if the thickness of the shock transition layer is larger than the electron gyroradius and less than the ion gyroradius. These drift instabilities compete with instabilities between the incoming upstream ions and the shock-reflected ions. It is unclear how the magnetic field and these electrostatic instabilities interplay and how this affects the width and the structure of the shock transition layer. More specifically, it is unclear if and how a magnetized shock can be identified on radiographic images. 

Here we examine by means of a particle-in-cell (PIC) simulation the formation phase of a shock in the presence of a perpendicular magnetic field. Two plasma clouds collide at the speed $\approx 9 \times 10^5$ m$\textrm{s}^{-1}$ at a boundary, which is orthogonal to the collision direction. Both clouds consist of spatially uniform electrons and ions, which have the charge-to-mass ratio of fully ionized atoms with equal numbers of neutrons and protons. We model Deuterium ions for reasons discussed below. The electron temperature is set to 2.7 keV. The ions of both clouds have a temperature of 1.2 keV and 120 eV, respectively. This accounts for the fact that the ions of the laser-ablated plasma usually have a temperature that is different from that of the ambient medium because their sources are different. The ratio between the electron plasma frequency and the electron gyrofrequency is set to 100.

The collision speed of $\approx 0.003c$ between both plasma clouds and the speed $\approx$ 600 km/s of the forward shock are well below their counterparts in Ref. \cite{Kato10}. They are representative for shocks enwrapping SNR blast shells during their late evolution stage, for example that at the southwest (SW) or northwest (NW) rims of the SNR RCW86 \cite{Ghavamian01,Helder11,Castro13}. These shocks expand into a medium of density $\approx 1 \, \textrm{cm}^{-3}$ and a magnetic field with an amplitude between 0.1-1 nT (ISM) and 8-14 nT (post-shock field). The large postshock value of the magnetic amplitude implies that cosmic-ray driven instabilities are at work \cite{Berezhko03,Bell04}. The ratio between the electron plasma frequency and the electron cyclotron frequency ranges from 300 (1 nT) to 30 (10 nT) and our ratio of 100 should be representative for the upstream region of these shocks. The electron temperature is about 5 times higher than that observed close to these shocks and comparable to that in laser-generated plasma.
  
Our study addresses three questions. Firstly, do shocks form for our initial conditions and are they maintained by magnetic or by electrostatic forces? Secondly, if shocks form, what is the structure and the width of their transition layer? Thirdly, what is the ion distribution in the downstream region?

Our results are as follows. A hybrid structure, which is a combination of an electrostatic shock and a double layer \cite{Hershkowitz81}, forms at the front of the cloud with the hot ions. This hybrid structure, which is mediated by planar electrostatic fields, has been observed experimentally \cite{Ahmed13}. The magnetic field is expelled from the interval with a high thermal pressure of the plasma and it accumulates in front of the hybrid structure. The magnetic amplitude remains too weak to influence the ion dynamics. The electrostatic layer that moves in the direction of the plasma cloud with the high ion temperature does not steepen into a shock and its electric fields remain low. The likely reason is that this structure is moving at a speed below the ion acoustic speed. The observation of only one shock is a direct consequence of our choice of different ion temperatures for both clouds. This implies that one may not always detect a shock doublet in a laser-plasma experiment, where the ions of the blast shell can have a different temperature than those of the ambient plasma. 

An ion acoustic instability develops ahead of the hybrid structure after a few tens of inverse ion plasma frequencies. These ion acoustic waves are driven by the instability between the incoming upstream ions and the ions that have been reflected by the hybrid structure or that leaked from the downstream into the upstream region. The relative speed of the counterstreaming ion beams exceeds the ion acoustic speed and these waves are thus oriented obliquely to the beam flow direction \cite{ForslundC}. The layer, in which we find strong electric fields, widens by a factor of 40 and the unipolar electric field of the hybrid structure is replaced by an ensemble of ion acoustic waves. The turbulence layer heats up the ions orthogonally to the shock plane, while the potential difference between the denser downstream plasma and the dilute upstream plasma thermalizes the ions along the shock normal direction. The conversion of the directed flow energy of the upstream ions into thermal energy of the downstream ions implies that the turbulence layer corresponds to the transition layer of a shock. The turbulence layer has only resulted in a partial thermalization of the ions by the time the simulation has finished. 

The low ion flow speeds imply that the current of the ion filaments, which sustain the turbulence layer, is small. The magnetic field amplitudes we observe are not sufficient to modify the ion dynamics during the simulation runtime. As we go to higher shock speeds, magnetic filamentation instabilities develop that thermalize the incoming upstream ions by diffusive shock acceleration \cite{Kato10,Stockem14}. Shocks, which are mediated by a spatially uniform magnetic field, require much stronger magnetic fields \cite{Chapman05,Scholer06}. 

Our simulation confirms the finding in Ref. \cite{Kato10} that the electrostatic shocks, which are characterized by a planar electric field pulse with a thickness that is comparable to an electron skin depth, are transient structures. Such shocks are frequently observed in laboratory plasma. Shocks mediated by electrostatic turbulence, which are more similar to the astrophysical shocks that have evolved over long times, take longer to form. Our simulation predicts a timescale of 20 ns or more for a shock, which develops in an ambient medium of density $10^{15} \mathrm{cm}^{-3}$. 

Our paper is subdivided in the following way. Section 2 discusses the equations, which are solved by a PIC code and the initial conditions of the simulations. Our results are presented in Section 3 and discussed in Section 4.

\section{The simulation code, the initial conditions and the experiment}  
 
\subsection{The particle-in-cell method}

PIC codes approximate a plasma by an ensemble of computational particles (CPs). Each CP $j$ of species $i$ has a position $\mathbf{x}_j$ and velocity $\mathbf{v}_j$. It has a charge-to-mass ratio $q_j / m_j$, which has to be equal to that of the species $i$, but the same does not necessarily have to hold for both values on their own. The ensemble of all CPs of the plasma species $i$ approximates its phase space density $f_i (\mathbf{x},\mathbf{v},t)$. The electromagnetic fields are updated via an approximation of Maxwell's equations on a grid. Most PIC codes evolve the fields through a discretized form of Amp\`ere's law and of Faraday's law. 
\begin{eqnarray}
\nabla \times \mathbf{B} = \mu_0 \mathbf{J} + \mu_0 \epsilon_0 \partial_t 
\mathbf{E}, \\
\nabla \times \mathbf{E} = -\partial_t \mathbf{B}.
\end{eqnarray}
Gauss' law is either fulfilled as a constraint or through a correction step while $\nabla \cdot \mathbf{B} = 0$ is usually preserved to round-off precision. The plasma is approximated by CPs, which correspond to Lagrangian markers, and the fields are updated on an Eulerian grid. Both components have to be connected through suitable interpolation schemes. 

The algorithm, with which an explicit PIC code advances the plasma in time, is the following: The charge density and the current density contributions of each CP are interpolated to the neighboring grid cells with the help of a shape function, which depends on the selected interpolation order. The macroscopic charge density $\rho (\mathbf{x},t)$ and the current density $\mathbf{J}(\mathbf{x},t)$ on the grid are obtained by summing up the interpolated microscopic contributions of all CPs of all species. The electromagnetic fields $\mathbf{E}(\mathbf{x},t)$ and $\mathbf{B}(\mathbf{x},t)$ are updated with $\mathbf{J}(\mathbf{x},t)$ and $\rho (\mathbf{x},t)$. The updated electromagnetic fields are interpolated to the position $\mathbf{x}_j$ of each CP and its momentum $\mathbf{p}_j = q_i \Gamma_j \mathbf{v}_j$ ($\Gamma_j$: relativistic factor) is updated through a discretized form of the relativistic Lorentz force equation $d\mathbf{p}_j / dt = q_j \left ( \mathbf{E}(\mathbf{x}_j) + \mathbf{v}_j \times \mathbf{B}(\mathbf{x}_j) \right )$. Each time this cycle is completed, the plasma is advanced in time by one time step $\Delta_t$. The PIC simulation method is discussed in more detail elsewhere \cite{Dawson83}. We use the EPOCH PIC code \cite{Cook11,Brady12}.

\subsection{The initial conditions of the simulation} 

The simulation plane is resolved by $10^4$ grid cells along $x$ and by 600 cells along $y$. The boundary conditions are open along $x$ and periodic along $y$. We introduce two plasma clouds, each consisting of electrons and ions. The ions have a charge-to-mass ratio that equals that of fully ionized atoms with equal numbers of protons and neutrons. We distribute the plasma as follows: We split the simulation box in two halves along $x$ and the system is uniform along $y$. We place one plasma cloud in the left half and one in the right half. Each cloud is spatially uniform and has identical charge density contributions from electrons and ions. Both plasma clouds are equally dense and the number densities of the electrons and of the ions are $n_0$, respectively. The equality of the number densities of both species implies that we model Deuterium ions. The electron plasma frequency is $\omega_{pe}={(n_0 e^2/m_e \epsilon_0)}^{0.5}$ and the ion plasma frequency is $\omega_{pi}\approx \omega_{pe}/60$. The electrons and the ions of each cloud have Maxwellian velocity distributions with an equal mean speed. The electrons of both clouds have the same temperature $T_e = 2.7$ keV. The ions of the left cloud have the temperature $T_{i,L}=1.2$ keV and those of the right cloud have the temperature $T_{i,R}=120$ eV.

Both clouds have the same mean speed modulus $v_c = 4.4 \times 10^5$ m$\textrm{s}^{-1}$ along $x$ and their mean speed along $y,z$ is set to zero. The right-moving cloud in the domain $x \le 0$ moves to increasing values of $x$ and the left-moving cloud in the domain $x > 0$ moves to decreasing values of $x$. Both plasma clouds touch at the start of our simulation and they thus interpenetrate immediately after the simulation has started. An initial spatial separation of both clouds along the x-direction would delay their collision. A delay and their spatial separation would imply that electrons can flow from both clouds into the vacuum that separates the ions. The resulting space charge would give rise to the formation of rarefaction waves at the front ends of both clouds and to a redistribution of the electromagnetic fields. 

We introduce a spatially uniform perpendicular magnetic field $B_{z,0}$ with the strength $\omega_{ce} /\omega_{pe} = 10^{-2}$ with $\omega_{ce}=eB_{z,0}/m_e$ and a convective electric field along $y$ with the modulus $|E_{y,0}| = v_c B_{z,0}$. The other field components are set to zero at the simulation's start. Each species is resolved by $6 \times 10^8$ CPs or 200 CPs per cell. The resolved ranges along $x,y$ are $-265 \le x/\lambda_e \le 265$ and $0 \le y / \lambda_e \le 32$, where $\lambda_e = c / \omega_{pe}$ is the electron skin depth. The simulation is evolved for a total time of $T_s \omega_{pi} = 491$ through $8.34 \times 10^5$ time steps of constant duration $\Delta_t \omega_{pi} = 5.9 \cdot 10^{-4}$. 

We have selected Deuterium ions for the following reason. Their charge-to-mass ratio equals that of fully ionized atoms, which are composed of equal numbers of protons and neutrons. Such ions typically form the ambient plasma and a substantial fraction of the blast shell plasma in the experiment. Their equal charge-to-mass ratio implies that these ions have the same ion plasma frequency $\omega_{pi}={(n_0 Z^2e^2/\epsilon_0 m_i)}^{1/2}$ if their total charge density $Z e n_0$ stays the same. The ion charge state is $Z$. Their ion cyclotron frequencies $\omega_{ci} = Ze B_{z,0} / m_i$ are equal as well. These ions also have the same ion skin depth $c/\omega_{pi}$, ion acoustic speed $\propto {(Z/m_i)}^{1/2}$ and Alfv\'en speed $\propto {(n_i m_i)}^{-1/2}$. The latter is true as long as the positive charge density is the same, which we exemplify as follows. The charge of doubly ionized Helium is twice that of Deuterium. Replacing Deuterium with He$^{2+}$ ions leaves the charge density unchanged if the ion number density $n_i$ is halved. This implies that $n_i m_i$ remains unchanged, because the mass of He$^{2+}$ is twice that of Deuterium. All characteristic plasma frequencies, the ion acoustic speed and the Alfv\'en speed do thus not depend on the particular choice of the ion species, as long as the charge-to-mass ratio is the same. The only plasma parameter that depends on the ion mass and not on the charge is the ion thermal speed $v_{ti} = {(k_B T_i / m_i)}^{1/2}$. Deuterium ions have the largest thermal speed for a given ion temperature $T_i$ and they are thus providing the strongest ion Landau damping of ion acoustic waves, as discussed in Chapter 4.2 in Ref. \cite{Treumann}. If the ion acoustic instability develops for Deuterium ions, then it will also occur for heavier ions with the same charge-to-mass ratio.

\subsection{The shock model}

\begin{figure}
\begin{center}
\includegraphics[width=10cm]{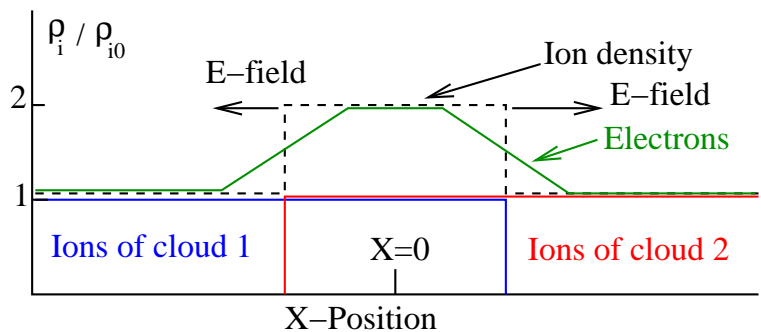}
\caption{The generation mechanism of electrostatic shocks: Two plasma clouds interpenetrate in the center of the simulation box and the total ionic charge density (dashed line) increases locally beyond that of each cloud. Electrons (green line) stream out of this cloud, leaving behind a positively charged overlap layer. This layer goes on a positive potential with respect to the incoming plasma clouds. The electric field will interact with the incoming plasma: its electrons are accelerated towards the overlap layer and the ions are slowed down. Some incoming ions are reflected.}\label{ShockModel}
\end{center}
\end{figure}

The collision of the plasma clouds in our simulation will result in a pile-up of ions, which is illustrated in Fig. \ref{ShockModel} under the assumption that the ions are cold and form a sharp front.
Their much higher mobility lets some electrons stream out of the ion overlap layer. Negatively charged sheaths develop just outside of the ion overlap layer and positively charged sheaths just inside of it. This space charge results in an electric field that puts the overlap layer on a positive potential compared to both surrounding plasma clouds. This potential traps a fraction of the electrons inside the overlap layer and it accelerates the electrons as they flow into the overlap layer. The potential develops on a time scale that is comparable to a few times the inverse electron plasma frequency.

The strength of the ambipolar electrostatic field depends only on the thermal pressure gradient of the electrons. A maximum speed thus exists up to which ions can be slowed down sufficiently to trigger the formation of a nonrelativistic unmagnetized shock, which is typically a few times the ion acoustic speed. The incoming ions are slowed down significantly in this case and some are reflected. This ion phase space structure is an electrostatic shock. The electric field also accelerates the ions that move to the boundary of the overlap layer, which form a double layer. A hybrid structure is one, in which an electrostatic shock and a double layer coexist \cite{Hershkowitz81}. A perpendicular magnetic field traps electrons and it can strengthen their confinement to the overlap layer. The magnetic field thus allows for larger differences between the electron's thermal pressures upstream and downstream of the shock. This implies that the maximum electrostatic potential can be increased by the magnetic field, which can stabilize shocks at larger speeds. This effect is negligible here due to our weak magnetic field. 

Larger collision speeds imply that the incoming ions do not lose enough kinetic energy as they move into the overlap layer and they thermalize via ion-ion beam instabilities \cite{Dieckmann13a} or via the Buneman instability \cite{Dieckmann06}. If instabilities can not thermalize the plasma, then the ions are reflected by the magnetic field on a time scale that is comparable to the inverse ion gyro-frequency \cite{Chapman05,Scholer06}. Filamentation instabilities will become important at ion beam speeds that exceed a few percent of $c$ \cite{Kato10,Stockem14}.

The relative speed between the counterstreaming ion populations along the collision direction is decreased to a value that is comparable to their thermal speed, if the kinetic energy of the ions in the rest frame of the overlap layer is sufficiently low. This slowdown takes place well behind the front of the interpenetrating ion beams. The counterstreaming ion beams have a similar mean speed along the collision direction in what we call the downstream region, while their mean speeds are close to their respective initial collision speed in the overlap region.

An electrostatic shock is characterized by a compression of the ions through their slowdown along the shock propagation direction. This slowdown does not affect the ion distribution in the orthogonal directions. The shock will generate a non-Maxwellian ion velocity distribution in the overlap layer in Fig. \ref{ShockModel}, which has a larger thermal velocity spread along $x$ than along $y$ and $z$ \cite{Dieckmann13a}. However, a plasma thermalization by a shock crossing, which yields heating to the same temperature in all directions, is assumed by the hydrodynamic or magnetohydrodynamic models invoked in astrophysical settings. These shocks are discontinuities that separate two plasmas with distinct macroscopic properties such as the flow speed, the temperature and the magnetization. 

\section{The simulation}

We discuss in what follows the plasma and in-plane electric field distributions at the times $t=$ 10.6, 53, 106 and 491. Time and space are expressed in units of $\omega_{pi}^{-1}$ and $\lambda_e = c / \omega_{pe}$. The ion density $\tilde{n}_i (x,y,t)$, the magnetic field energy density $\tilde{E}_B (x,y,t) = \mathbf{B}^2 (x,y,t) / 2\mu_0$ and the electric field energy density $\tilde{E}_E (x,y,t) = \epsilon_0 \mathbf{E}^2 (x,y,t) / 2$ are used to track the plasma evolution. These quantities are averaged along $y$ over the box length $L_y = 32 \lambda_e$ giving $\tilde{n}_i (x,t)=L_y^{-1} \int \tilde{n}_i (x,y,t) \, dy$, $\tilde{E}_E (x,t) = (\epsilon_0 / 2L_y) \int \mathbf{E}^2 (x,y,t) \, dy$ and $\tilde{E}_B (x,t) = {(2\mu_0L_y)}^{-1}\int \mathbf{B}^2 (x,y,t) \, dy$. The normalized ion density is $n_i (x,t) = \tilde{n}_i (x,t) / \tilde{n}_i (x,t=0)$. The field energy densities are normalized as $E_B (x,t) = 2\mu_0 \tilde{E}_B (x,t) / B_{z,0}^2$ and $E_E (x,t) = 2\mu_0 \tilde{E}_E (x,t) /B_{z,0}^2$.

\begin{figure}
\begin{center}
\includegraphics[height=5.8cm]{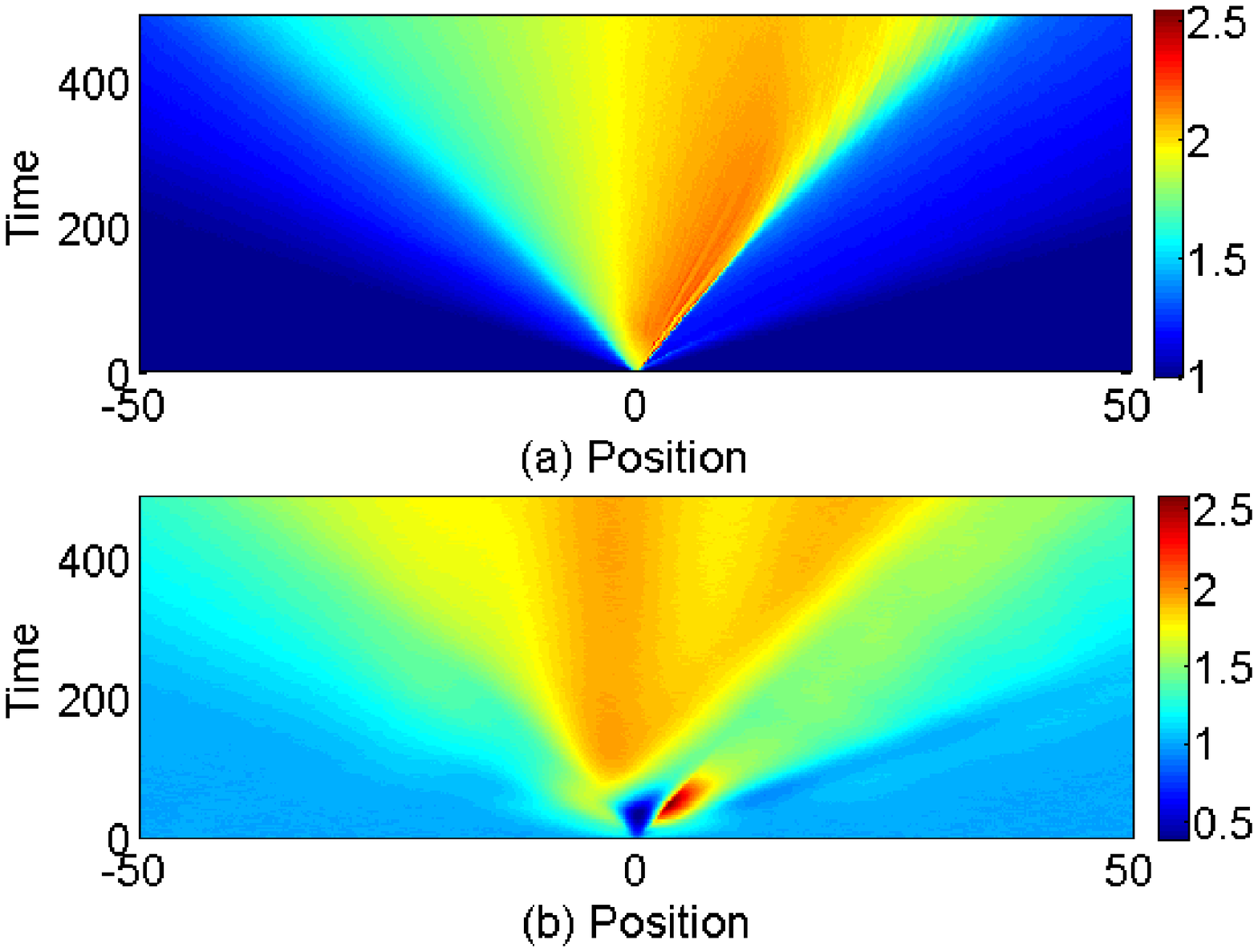}
\includegraphics[height=5.8cm]{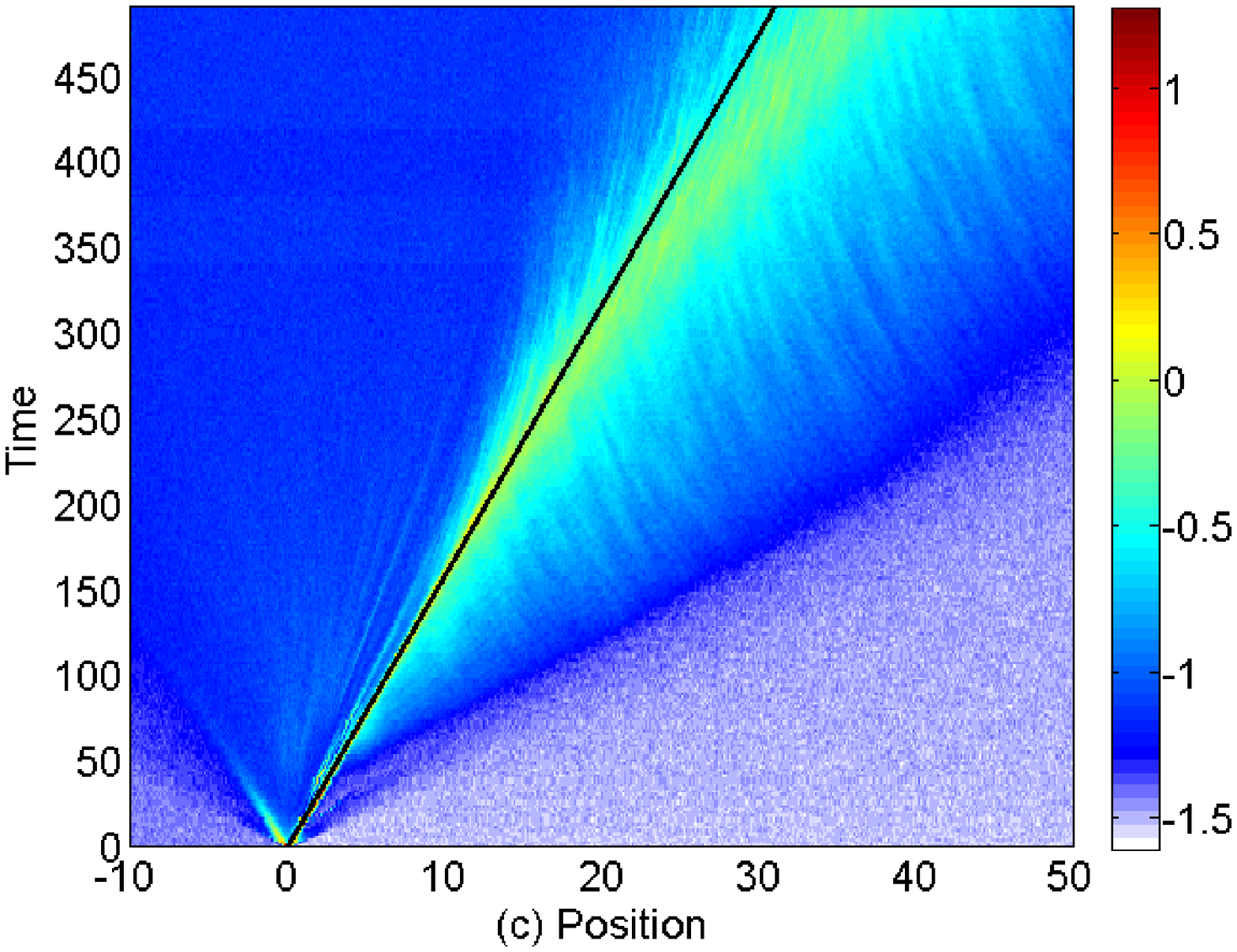}
\caption{The ion density $n_{i}(x,t)$ is displayed in panel (a) and panel (b) shows $E^{1/2}_B(x,t)$ on a linear color scale. Panel (c) displays $E_E (x,t)$ on a 10-logarithmic scale. The overplotted black line corresponds to a speed of $v_s = 1.3 \times 10^5$ m$\textrm{s}^{-1}$.}
\label{Average}
\end{center}
\end{figure}

Figure \ref{Average} displays their spatio-temporal evolution.
Figure \ref{Average}(a) reveals a central region with $n_i (x,t) > 1.5$, which is expanding at a constant speed in both directions. Its front reaches $|x| \approx 30$ at $t=491$. The ion structure is not symmetric with respect to $x=0$. The peak ion density $n_i (x,t) \approx 2.5$ is reached in the interval $x>0$, where we also observe the steepest ion density gradients. A density plateau with $n_i (x,t) \approx 1.7$ is present in the interval $-25 < x < -10$ and the density gradually decreases to 1.3 within $-35 < x < -25$ at $t$=491. We thus expect that the plasma distribution at the structure, which is moving to increasing values of $x>0$, differs from that at the structure that moves to decreasing values of $x<0$. Fast structures with $n_i (x,t) \approx 1.3$ can be seen, which cross the edge of the displayed spatial interval at $t\approx 200$.

The magnetic field energy density is shown in Fig. \ref{Average}(b). The magnetic field distribution is practically uniform along $y$ during the entire simulation time (not shown) and $E_B^{1/2}(x,t)$ thus expresses the magnetic field amplitude in units of $B_{z,0}$. A short-lived bipolar magnetic structure with a peak amplitude of $E_B \approx 6$ is visible at $t < 50$ and $x\approx 0$. Thereafter, a more stable magnetic field distribution develops. A peak value of $E_B \approx 4$ is observed at $x \approx 0$ after $t\approx 50$. The magnetic front that reaches $x\approx 30$ at $t=491$ is correlated with the front of the high density region and we also observe elevated magnetic field energy densities within the fast ion density structures. 

A weak pulse is present in $E_E (x,t)$ at early times in Fig. \ref{Average}(c) and in the interval $x<0$. A strong and sharp electric field pulse is observed in the interval $x>0$ until $t\approx 40$. The pulse propagates to increasing values of $x$ at the same speed $\approx 1.3 \times 10^5$ m$\textrm{s}^{-1}$ as the location of the steepest gradient in the ion density in Fig. \ref{Average}(a). The electric field pulse broadens in time and it covers an interval along $x$ with a width of about $10\lambda_e$ at $t=491$. Weaker electrostatic fields cover an even wider interval at this time. The initial concurrence between the electric field pulse in the interval $x>0$ and the location with the steepest ion density gradient suggests that, at least until $t \approx$ 40, the pulse corresponds to the ambipolar electric field, which is a consequence of the electron's thermal pressure gradient. The electric field pulse thus characterizes the location of an electrostatic shock, of a double layer or of their combination. The magnetic pressure gradient $\propto d_x \mathbf{B}^2 (x,t)$ appears to be too weak to drive an electrostatic field, since there is no visible correlation between the steepest spatial gradients of $E_B(x,t)$ and the distribution of $E_E(x,t)$. 

The Mach number of the electrostatic pulse that moves to increasing values of $x>0$ is the following. The Alfv\'en speed is $v_A \approx 5 \times 10^4$ m$\textrm{s}^{-1}$. The ion acoustic speed is $c_s = {(\gamma_c k_B (T_e + T_i) Z / m_i)}^{1/2} =  4.75 \times 10^5$ m$\textrm{s}^{-1}$ for the adiabatic constant $\gamma_c = 5/3$, which we take for simplicity to be the same for the electrons and the ions, and for values of the ion charge $Z$ and mass $m_i$ that correspond to those of our ions. The ion acoustic speed is 20$\%$ higher in the right-moving cloud due to the hotter ions. Given the high plasma $\beta \approx n_0 k_B (T_e + T_i) / (B^2_{z,0}/2 \mu_0) > 10^2$ and $c_s \gg v_A$, the magnetosonic modes have dispersive properties that can not be distinguished from those of an ion acoustic wave. Magnetosonic waves can also not develop because the simulation time $t=491$ resolves only 8\% of one ion gyro-orbit. If a shock forms during the simulation time, it must be electrostatic. The pulse speed $\approx 1.3 \times 10^5$ m$\textrm{s}^{-1}$ corresponds to $\approx 0.27 c_s$ in the simulation frame and to $1.2 c_s$ in the reference frame of the left-moving cloud. The Mach number of the pulse in the right-moving plasma with its hotter ions may be below unity, explaining the asymmetry between the intervals $x<0$ and $x>0$ in Fig. \ref{Average}. 

\subsection{Time 10.6: The electrostatic shock / double layer hybrid structure} 

\begin{figure}
\begin{center}
\includegraphics[width=10cm]{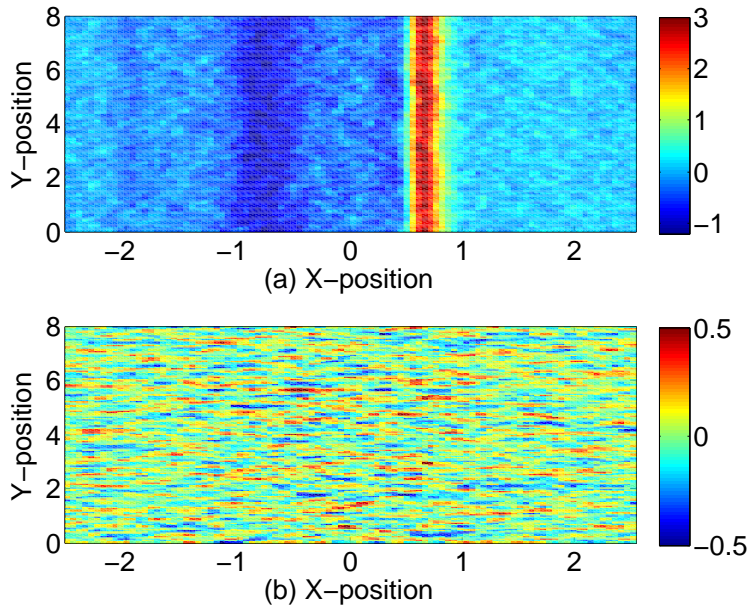}
\caption{The in-plane electric field in units of $10^{-2} m_e c \omega_{pe} / e$: Panel (a) shows $E_x$ and panel (b) shows $E_y$ in a sub-interval of the simulation box. The time is $t = 10.6$.}\label{electric10.6}
\end{center}
\end{figure}

Figure \ref{electric10.6} shows that the electric field is planar at this time and that it points along the plasma flow direction. The distribution of $E_x (x,y)$ shows a strong peak at $x\approx 0.7$ with a peak amplitude of $\approx 0.03$ and a width of $0.25 \lambda_e$. A second planar electric field distribution is present at $x\approx -0.9$ in Fig. \ref{electric10.6}(a). It spans a wider x-interval and it reaches a minimum value of $\approx -0.01$. The electric field polarization is such that the region enclosed by both pulses is on a higher potential than the plasma that surrounds them. Figure \ref{electric10.6}(b) shows only noise. This electric field configuration resembles the one observed in the experiment discussed in Ref. \cite{Ahmed13}.
 
\begin{figure}
\begin{center}
\includegraphics[height=5.7cm]{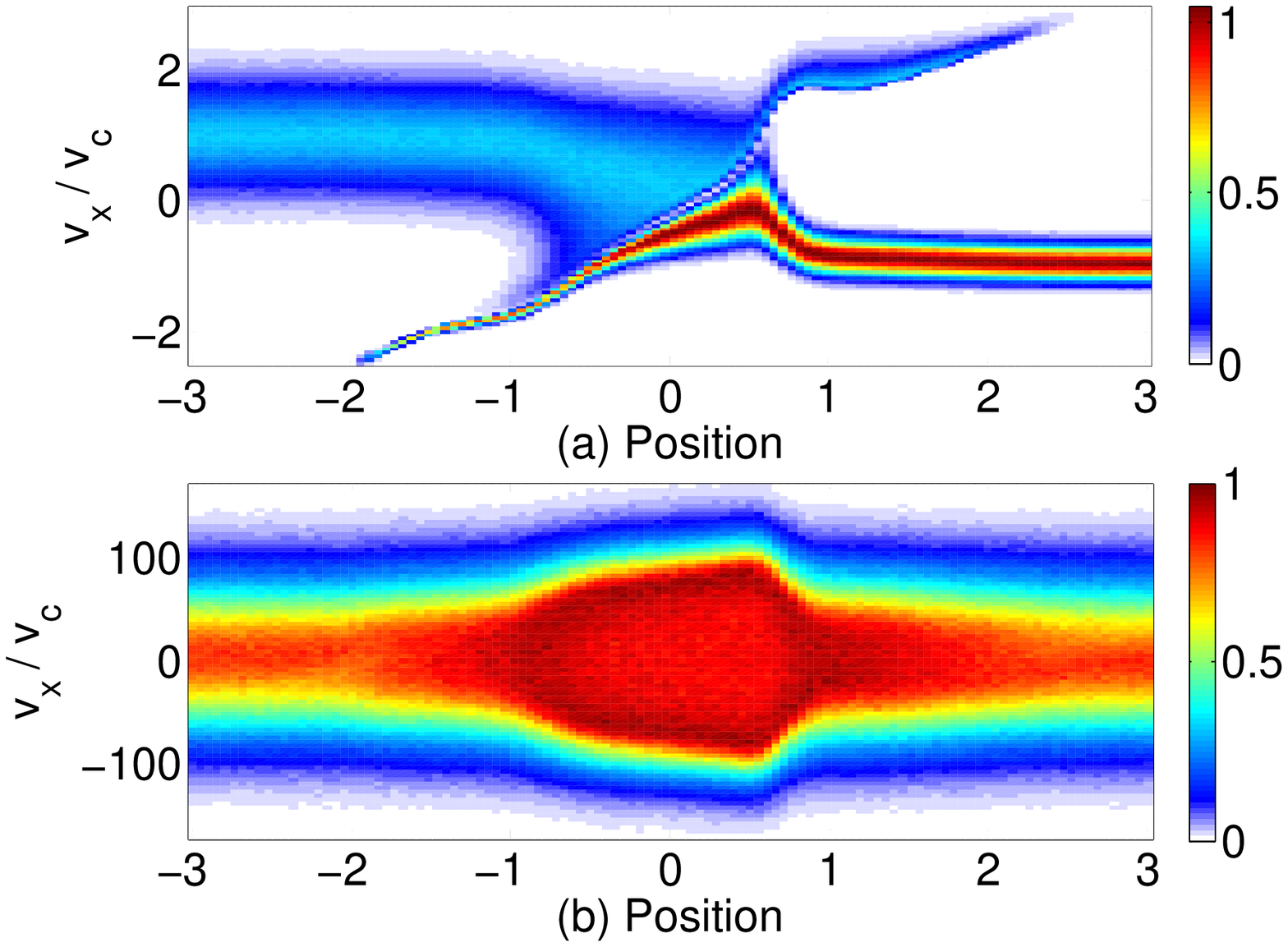}
\includegraphics[height=5.7cm]{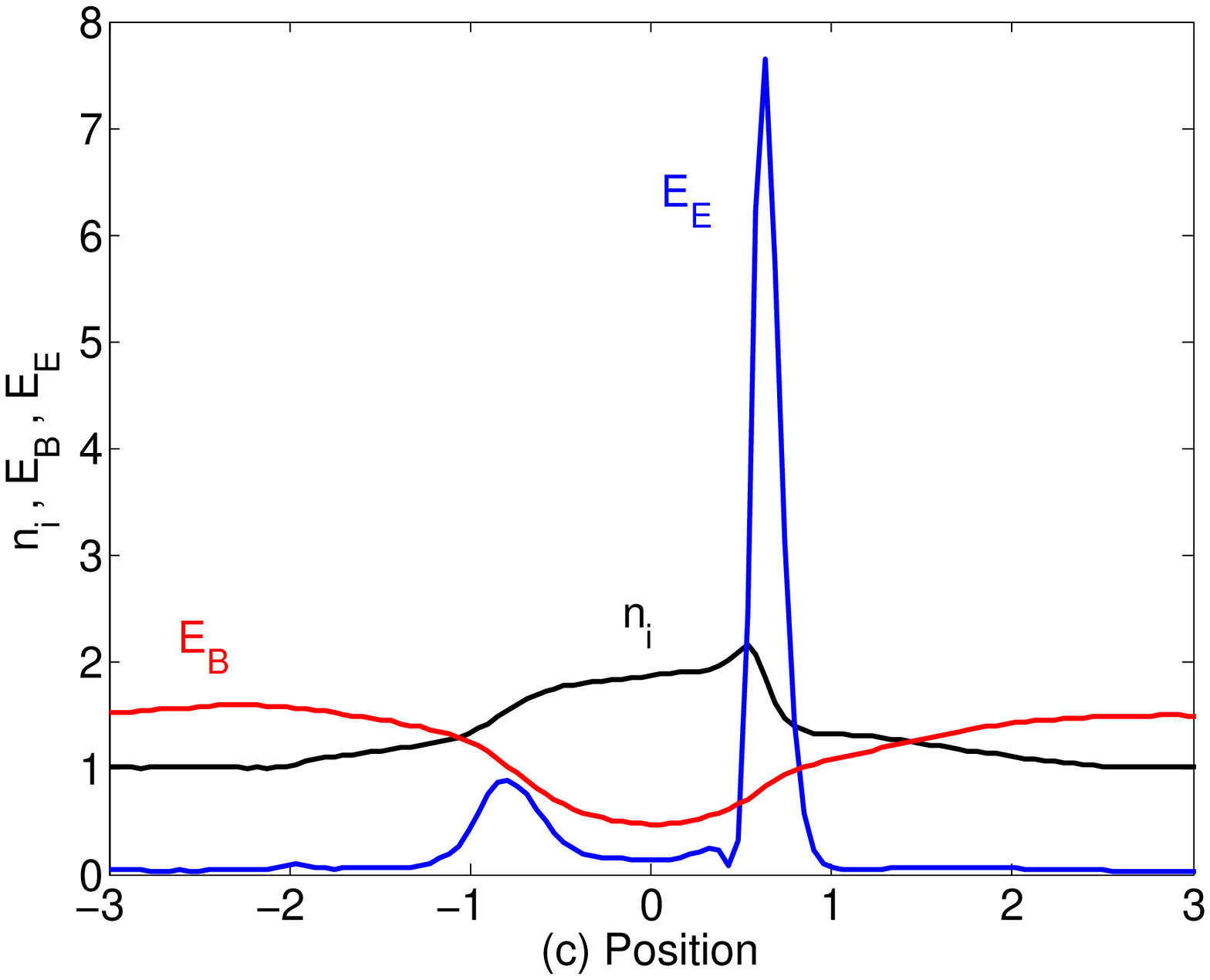}
\caption{Panel (a) shows the y-averaged ion phase space density distribution $f_i (x,v_x)$ and panel (b) the y-averaged electron phase space density distribution $f_e(x,v_x)$. The color scale is linear. Panel (c) shows $n_i (x,t)$ (black curve), the y-averaged magnetic field energy density $E_B(x,t)$ (red curve) and the electric field energy density $E_E(x,t)$ (blue curve). The simulation time is $t = 10.6$.}\label{phasespace10.6}
\end{center}
\end{figure}

The ion and electron phase space density distributions $f_e (x,v_x)$ and $f_i (x,v_x)$, which have been integrated along $y$, are shown in the Figs. \ref{phasespace10.6}(a,b). The phase space density distribution of the ions reveals overlap layers in the intervals $-2 < x < -1$ and $0.7 < x < 2.5$. The ion distributions outside of this interval show a single beam with a Maxwellian velocity distribution. The counterstreaming ion populations have merged along $v_x$ in the interval $-1<x<1$ to form the downstream region. The strong $E_x$ fields in Fig. \ref{electric10.6} have slowed down the ion beams to a degree that has let them merge along the $v_x$-direction. The ions of the left-moving cool ion beam are slowed down more and on a smaller spatial range, which is a consequence of the asymmetric distribution of $E_x$ in Fig. \ref{electric10.6}(a). We observe dilute ion beams in the intervals $1<|x|<2.5$. Their main source at this time are the ions that have crossed the downstream region and are accelerated by the ambipolar electrostatic field as they move into the overlap layer. This is a double layer. The incoming ions, which are slowed down as they move from the overlap layer to the downstream region, constitute an electrostatic shock if their speed change exceeds the ion acoustic speed. The ion phase space structure at $x\approx 0.7$ is thus a hybrid structure and, possibly, the one at $x \approx -1$. The differences between both plasma structures is a consequence of the different ion temperatures in both clouds. Both structures would be similar for equal ion temperatures.

The electron distribution in Fig. \ref{phasespace10.6}(b) shows a velocity distribution outside of the interval $-2.5 < x < 2.5$, which is close to the initial one. Hot electrons from within the downstream region leak into the overlap layer and some propagate upstream of the overlap layer. Their current is compensated by a return current and electrons are accelerated towards the shock. The velocity spread of the electrons and, thus, their thermal energy is largest close to the right-moving shock at $x\approx 0.7$ and it decreases rapidly with increasing values of $x$. The high thermal pressure gradient of the electrons yields the large electric field at $x \approx 0.7$. The electron phase space density shows a ring distribution within $-1< x < 1$ and a local minimum at $x\approx 0$ and $v_x \approx 0$. 

Figure \ref{phasespace10.6}(c) compares the ion density with the electric and magnetic field energy densities. The ion density reaches its peak value $n_i \approx 2$ at $x\approx 0.5$ and it decreases to $n_i \approx 1.3$ at $x\approx 0.9$. The energy density of the electric field shows its peak value in the interval $0.5 < x < 0.9$, confirming that its source is the electron thermal pressure gradient maintained by the ion density variation. The value of $E_E$ is elevated in the interval $-0.4 < x < 0.4$ and it shows a weak local maximum at $x\approx 0.3$ that is supported by a local positive ion density gradient. Another peak of $E_E$ is located within $-1.2 < x < -0.5$ and coincides again with an ion density gradient. The magnetic field energy density $E_B$ has a minimum value of $0.5$ at $x\approx 0$ and it increases to about $1.5$ at $|x| \approx 2.5$. It converges to $E_B = 1$ outside of the displayed interval. We attribute the depletion of $E_B$ at $x\approx 0$ to the electron's diamagnetic current $\mathbf{J}_M = -(\nabla p \times \mathbf{B})/B^2$. Its effect via Amp\`ere's law is to expel the magnetic field from regions with a high thermal pressure of the plasma. This magnetic expulsion can be observed experimentally \cite{Niemann13}.

\subsection{Time 53: The drift instability} 

\begin{figure}
\begin{center}
\includegraphics[width=10cm]{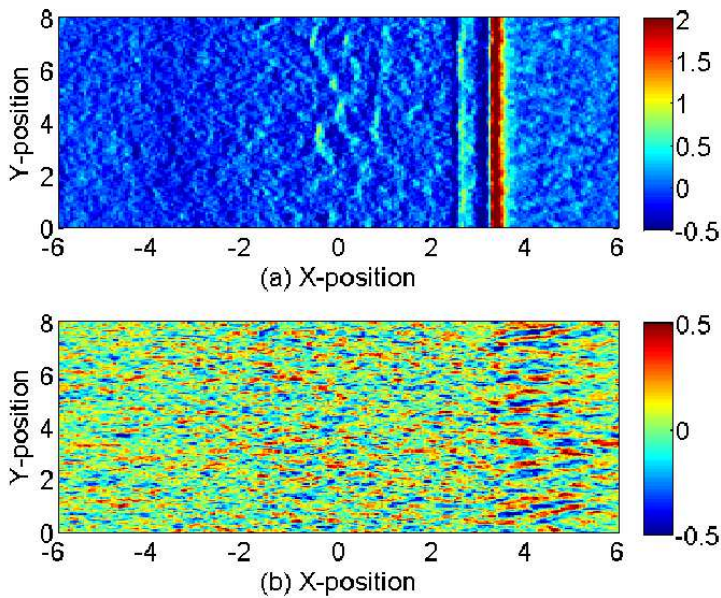}
\caption{The in-plane electric field in units of $10^{-2} m_e c \omega_{pe} / e$: Panel (a) shows $E_x$ and panel (b) shows $E_y$ in a sub-interval of the simulation box. The color scale is linear and $t = 53$.}\label{electric30}
\end{center}
\end{figure}

The electric field distribution in Fig. \ref{electric30} shows some differences compared to that at the earlier time. A tripolar planar pulse is centered at $x\approx 3$ in Fig. \ref{electric30}(a) and the strongest peak is located at $x\approx 3.5$. Weak wave structures are present in the interval $-1 < x < 2$ with a length of 1-2 $\lambda_e$ along $y$ and with an amplitude and width along $x$, which are comparable to those of the planar field structure at $x \approx$ 2.5. Structures with a wavelength of $\lambda \approx 0.5$ are visible in the interval $x>4$ in Fig. \ref{electric30}(b), which have no counterpart in $E_x$ and in $B_z$ (not shown). 

\begin{figure}
\begin{center}
\includegraphics[height=5.7cm]{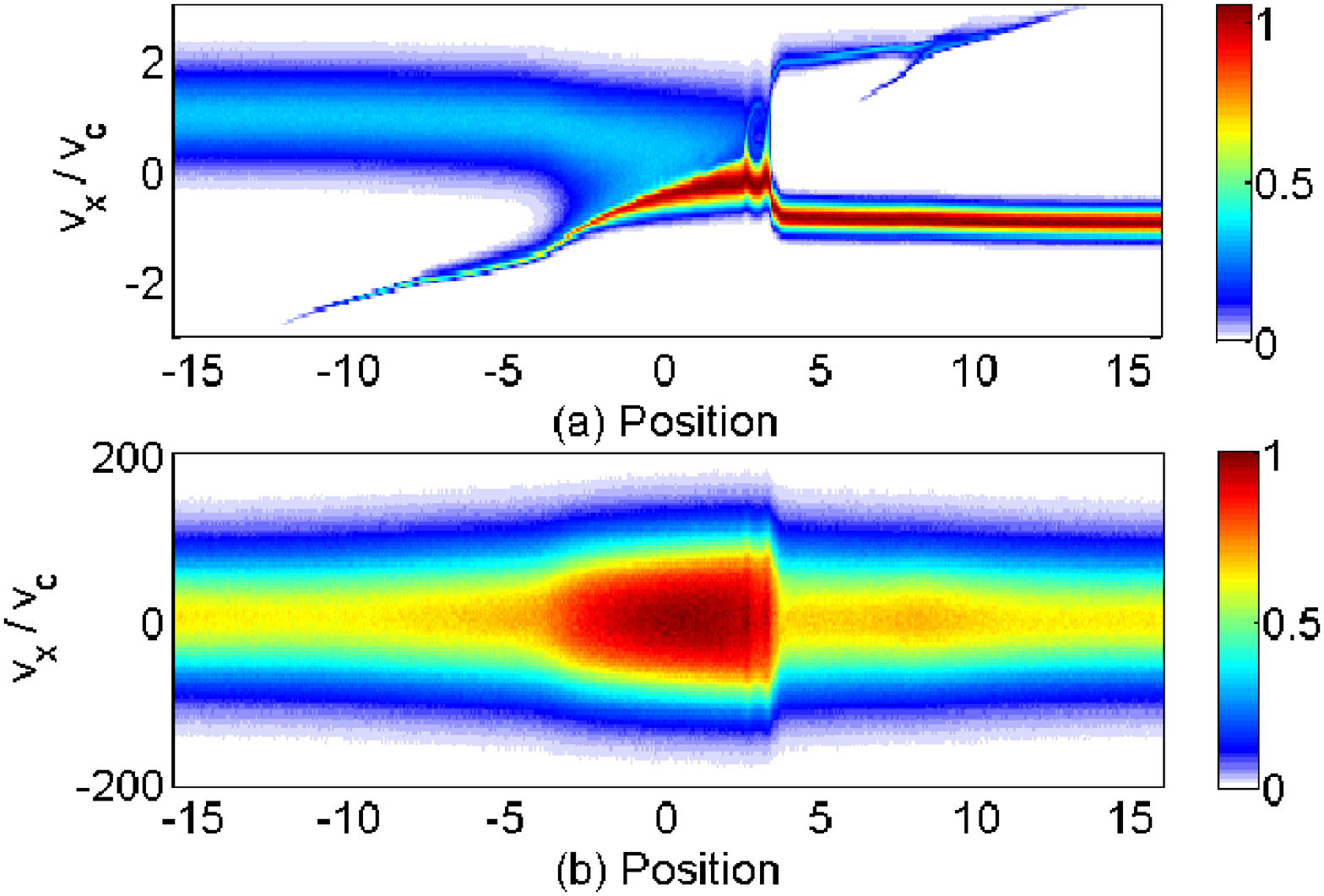}
\includegraphics[height=5.7cm]{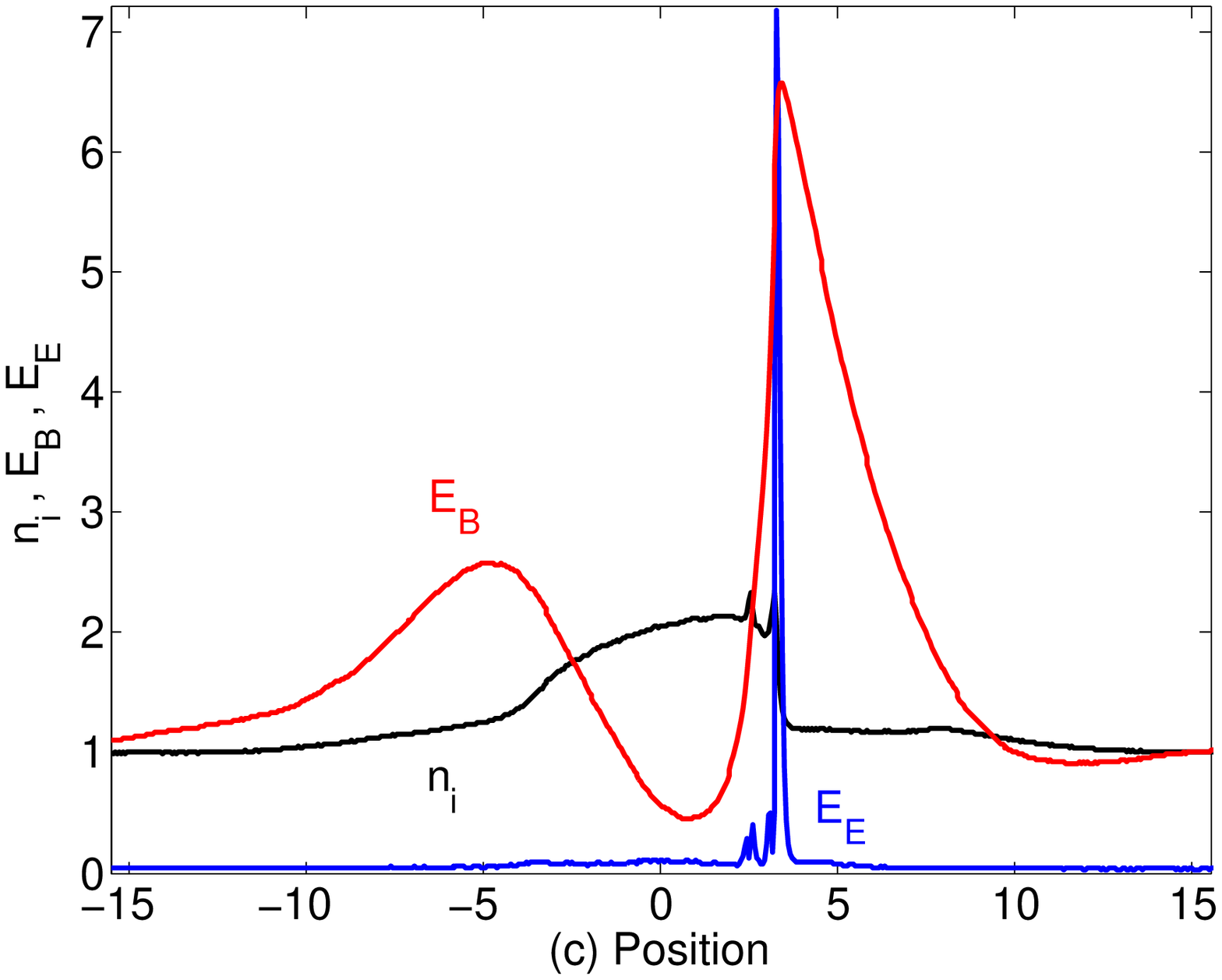}
\caption{Panel (a) shows the y-averaged ion phase space density distribution $f_i (x,v_x)$ and panel (b) the y-averaged electron phase space density distribution $f_e(x,v_x)$. The color scale is linear. Panel (c) shows $n_i (x,t)$ (black curve), the y-averaged magnetic field energy density $E_B(x,t)$ (red curve) and the electric field energy density $E_E(x,t)$ (blue curve). The simulation time is $t = 53$.}\label{phasespace53}
\end{center}
\end{figure}

The ion and electron phase space density distributions in Fig. \ref{phasespace53} reveal a hybrid structure at $x\approx 3.5$ with a transition layer thickness that is identical to that at $t$=10.6. The transition layer in the interval $x<0$, across which the mean speed of the ions changes from $v_c$ at $x=-15$ to the downstream value, is much wider. The ion distribution resembles that of a rarefaction wave that expands into an ambient plasma prior to the formation of a shock \cite{Sarri11}, which suggests that the overlap layer in the interval $x<0$ propagates at a speed that is below the ion acoustic speed. The electrons show a velocity distribution in the interval $-3< x < 3$, which is close to a Maxwellian distribution with a maximum at $v_x \approx 0$. The tip of the dilute ion beam in the interval $x>3.5$ and $v_x > 0$ and the tip of the ion beam in the interval $x<-5$ and $v_x < 0$ have moved away from $x=0$ by a distance of $\approx 13 \lambda_e$ at the time $t=53$. The fast ion density structures with $n_i (x,t) \approx 1.3$ in Fig. \ref{Average}(b) thus outline the overlap layer. The ion density reaches its maximum value within the downstream region.   

The ion density and the field energy densities in Fig. \ref{phasespace53}(c) show that the anti-correlation between the plasma density and the magnetic field energy density has strengthened. This anti-correlation is typical for a perturbation, which is propagating in the slow magnetosonic mode. The magnetic field amplitude reaches its peak value $\approx 2.5 B_{z,0}$ or $E_B \approx 6.5$ at $x\approx 4$ in Fig. \ref{phasespace53}(c) and it has steepened significantly at this location. The slow magnetosonic wave is linearly undamped for a propagation direction perpendicular to the B-field \cite{Barnes66}, but Fig. \ref{Average}(b) shows that the pulse is evanescent. Drift instabilities can limit the steepening of shock waves \cite{Forslund72}. They develop when electrons are accelerated along the shock boundary by gradient drifts, provided that the relative drift speed between electrons and ions exceeds the threshold of instability. Their thermalization yields non-linear damping. The waves in Fig. \ref{electric30}(b) have a wavevector, which points along the shock boundary. This polarization is typical for the electrostatic waves that result from drift instabilities. It has been proposed that the relative motion of electrons and ions, which triggers drift instabilities, is enforced by the $\mathbf{E}\times \mathbf{B}$-drift \cite{Forslund72}. This mechanism can, however, not be responsible for the drift instability in our simulation. The structures in $E_y$ develop in a broad x-interval ahead of the narrow pulse in Fig. \ref{electric30}(a). 

Plasma particles can also drift in a magnetic field gradient. The magnetic field is strong and it has steepened significantly within the interval $3.5 < x < 6$ at the time $t=53$ in Fig. \ref{phasespace53}(c). The electron gyroradius $r_c = v_{te}/(\omega_{ce} \lambda_e) \approx 0.73$ is smaller than the width of this interval and the net flow speed $v_c$ of the electrons is much less than the electron's thermal speed $v_{te}$. The electrons stay for a long time within the region with the large magnetic field gradient. The guiding center theory is applicable for the electrons, while the much heavier ions behave as if they were unmagnetized. The grad-B drift speed of electrons is given by $v_D = -(m_e v_{te}^2 / 2e B_z^2) \partial_x B_z$, since $B_x,B_y \approx 0$ and because $B_z$ varies only along $x$. 

\begin{figure}
\begin{center}
\includegraphics[width=10cm]{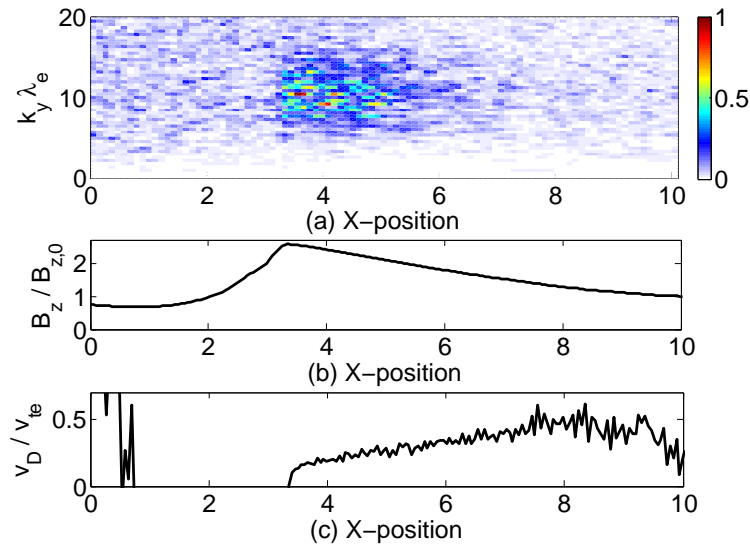}
\caption{Panel (a) shows the power spectrum $E^2_y (x,k_y)$ of the electric field $E_y$ close to the forward shock at $x\approx 3.5$. The power spectrum is normalized to its peak value and the color scale is linear. Panel (b) shows the y-averaged $B_z (x,t) / B_{z,0}$ distribution and panel (c) the normalized drift speed computed from the grad-B drift.}\label{GradB}
\end{center}
\end{figure}

The wave number and spatial distribution of the waves in Fig. \ref{electric30}(b) can be determined more accurately by taking the Fourier transform of $E_y (x,y,t=53)$ along $y$ and by computing its power spectrum ${|E_y (x,k_y,t=53)|}^2$. Figure \ref{GradB} compares it to $B_z (x,t=53)$, which has been averaged along $y$, and to $v_D {v}_{te}^{-1}$.
Most of the wave power is concentrated in the region $3.5 < x < 5.5$ and $6 < k_y \lambda_e < 15$ in Fig. \ref{GradB}(a). The power spectrum in Fig. \ref{GradB}(a) peaks at $k_y \lambda_e \approx 10$ equalling a wave length of $\lambda_u \approx 0.6 \lambda_e$. The electron drift speed along $y$ exceeds $0.2 v_{te}$ ahead of $x \approx 3.5$, which can destabilize the ion acoustic instability between electrons and ions \cite{Jackson60} and the electron cyclotron drift instability \cite{Forslund72}. Both instabilities yield electrostatic waves with a wavevector that is aligned with the drift speed. The waves grow in a spatial interval with a relatively high drift speed and with a high magnetic field amplitude, which points at the electron cyclotron drift instability as the responsible process. The growth rate of these waves can be a significant fraction of $\omega_{ce}$ and their wave numbers $k_y r_c$ can exceed unity. These waves can thus account for the oscillations with the short wave length $\lambda_u \approx 0.6 \lambda_e$ or $2\pi \lambda_u^{-1} r_c \approx 7$ in Fig. \ref{GradB}(a). The magnetic field gradient reverses its sign as we go to $x<3.5$ but it is still relatively large behind the hybrid structure. We would expect that a drift instability develops behind $x\approx 3.5$ too. The wave growth might be delayed or suppressed by the higher electron temperature in the downstream region. 

\begin{figure}
\begin{center}
\includegraphics[width=10cm]{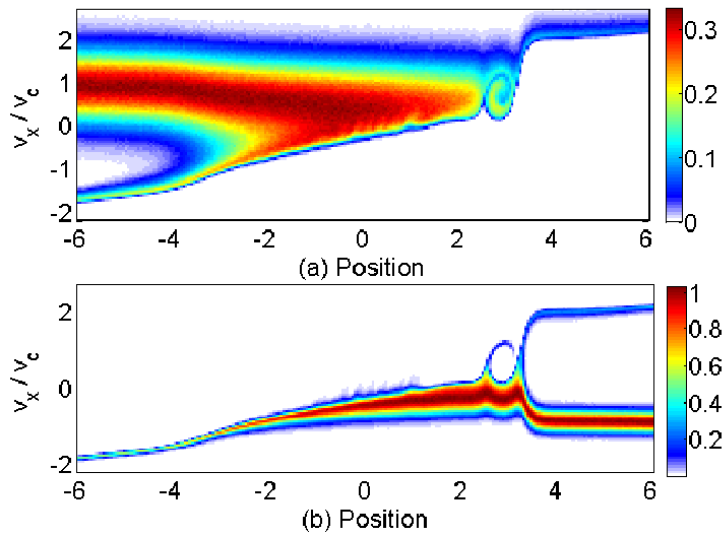}
\caption{Panel (a) shows the y-averaged ion phase space density distribution $f_i (x,v_x)$ of the left ion beam and panel (b) shows that of the right ion beam. The color scale is linear and $t = 53$.}\label{IonHole}
\end{center}
\end{figure}

The electric field energy density $E_E (x,t)$ shows a strong peak at $x\approx 3.5$ where we find the steepest ion density gradient and it is thus caused by the hybrid structure. The electric field energy density shows two local maxima at $x\approx 2.5$ and at $x\approx 3$, which reflects the tripolar nature of the electric field at $x\approx 3$ in Fig. \ref{electric30}(a). The reason for why we get a tripolar pulse rather than an unipolar pulse can be determined through a separation of the phase space density distributions of both counterstreaming ion beams. Figure \ref{IonHole} (a) shows the distribution of the right-moving ion beam and Fig. \ref{IonHole} (b) that of the left-moving ion beam. 
The ion distributions demonstrate unambiguously that the structure at $x=3.5$ is not a pure electrostatic shock, because some of the ions in Fig. \ref{IonHole}(a) cross this position and are accelerated to larger $x$. The latter ion structure corresponds to a double layer. The incoming ions in Fig. \ref{IonHole}(b) with $x > 3.5$ and $v_x < 0$ are slowed down as they approach the hybrid structure. Most ions cross this position and some are reflected. This distribution is that of an electrostatic shock. A vortex, which is also known as ion phase space hole, is present in both beams in the interval $2.5 < x < 3.5$ and $0 < v_x / v_c < 1.2$. An ion phase space hole \cite{Schamel86,Eliasson06} corresponds to a local excess of negative charge and is thus characterized by a bipolar electric field pulse. Ion phase space holes are stable if the electron temperature is much larger than the ion temperature and they are increasingly damped as the temperatures equilibrate. The ion phase space hole is responsible for two of the three electric field peaks in Fig. \ref{electric30}(a). A combination of a unipolar electrostatic field pulse and additional field oscillations has been observed in the experiment discussed in Ref. \cite{Romagnani08}, which attributed this wave train to a shock and to solitons. 

\subsection{Time 106: The onset of the ion-ion instability} 

\begin{figure}
\begin{center}
\includegraphics[width=10cm]{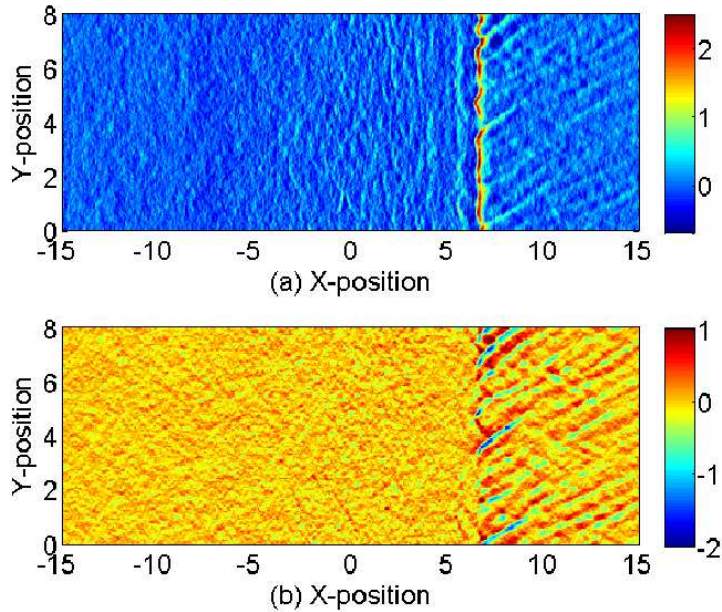}
\caption{The in-plane electric field in units of $10^{-2} m_e c \omega_{pe} / e$: Panel (a) shows $E_x$ and panel (b) shows $E_y$ in a sub-interval of the simulation box. The color scale is linear and $t = 106$.}\label{electric60}
\end{center}
\end{figure}

The in-plane electric field distribution at $t=106$ is shown in Fig. \ref{electric60}. A strong quasi-planar electric field pulse is located at $x\approx 6.5$ in Fig. \ref{electric60}(a) whose amplitude is modulated along $y$. A second weaker quasi-planar electric field pulse is trailing it at $x\approx 5.5$. This weaker pulse is still caused by an ion phase space hole. Localized electric field patches in the interval $0 < x < 5$ have an extent $\approx \lambda_e$ along $y$. Oblique wave structures have developed ahead of $x\approx 6.5$ in Fig. \ref{electric60}(a,b). Such oblique wave structures are driven by unmagnetized ion beams that move relative to each other at a speed that exceeds the ion acoustic speed \cite{ForslundC}. 
Only noise is observed in Figs. \ref{electric60}(a,b) for $x<-6$. 

The growth of the oblique structures coincides with the development of the drift instability, which can be seen from the supplementary movie 1. This movie animates the time-evolution of both in-plane electric field components close to the hybrid structure for $0 < t < 491$. The electric field components are expressed in units of $10^{-2} m_e c \omega_{pe} / e$. It is unclear if the drift instability, which results in waves with wavevectors $k_y$ that are similar to those of the ion acoustic wave, provides the seed for the oblique waves or if the development of the ion acoustic instability is modified by the drift current.

\begin{figure}
\begin{center}
\includegraphics[height=5.7cm]{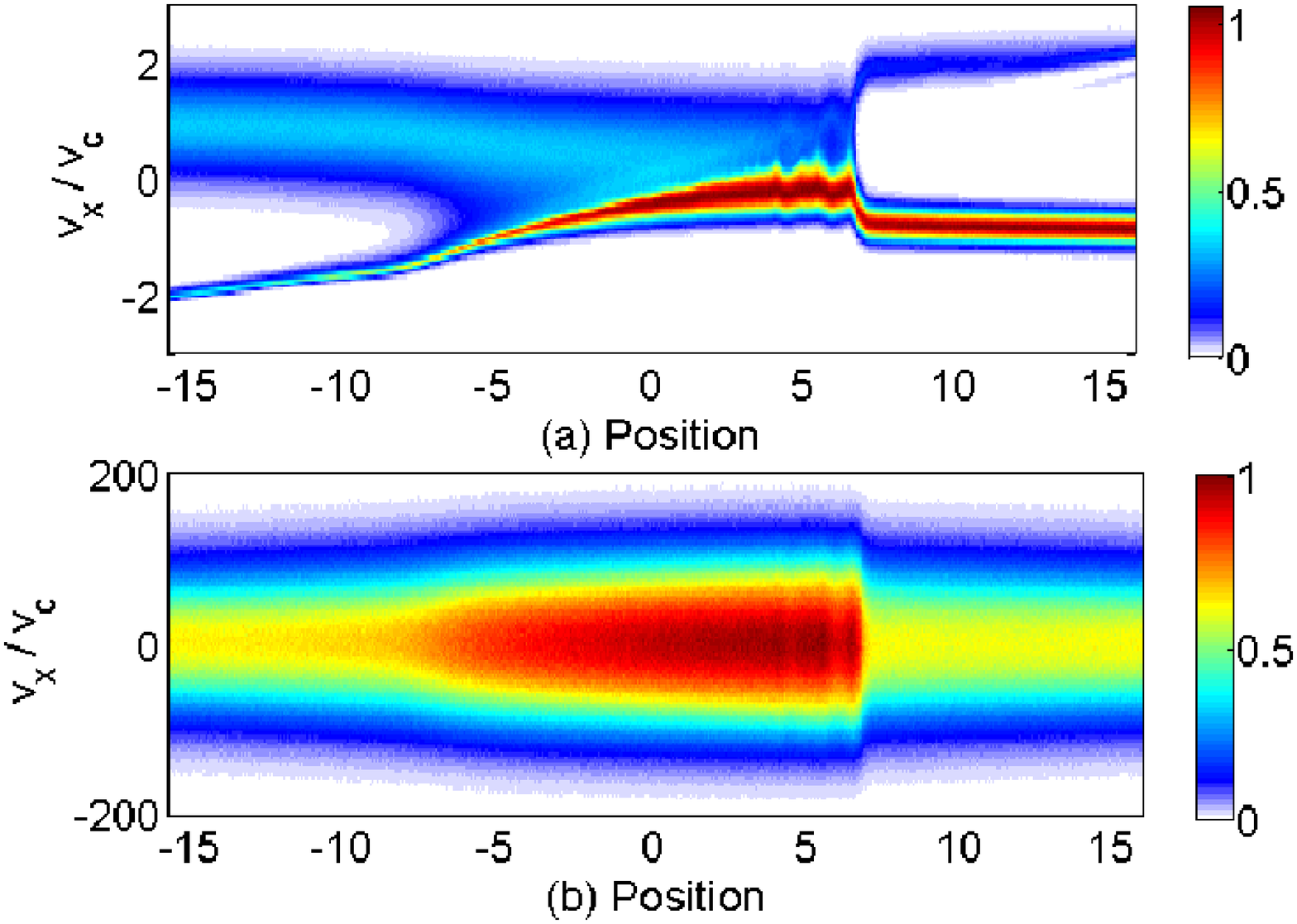}
\includegraphics[height=5.7cm]{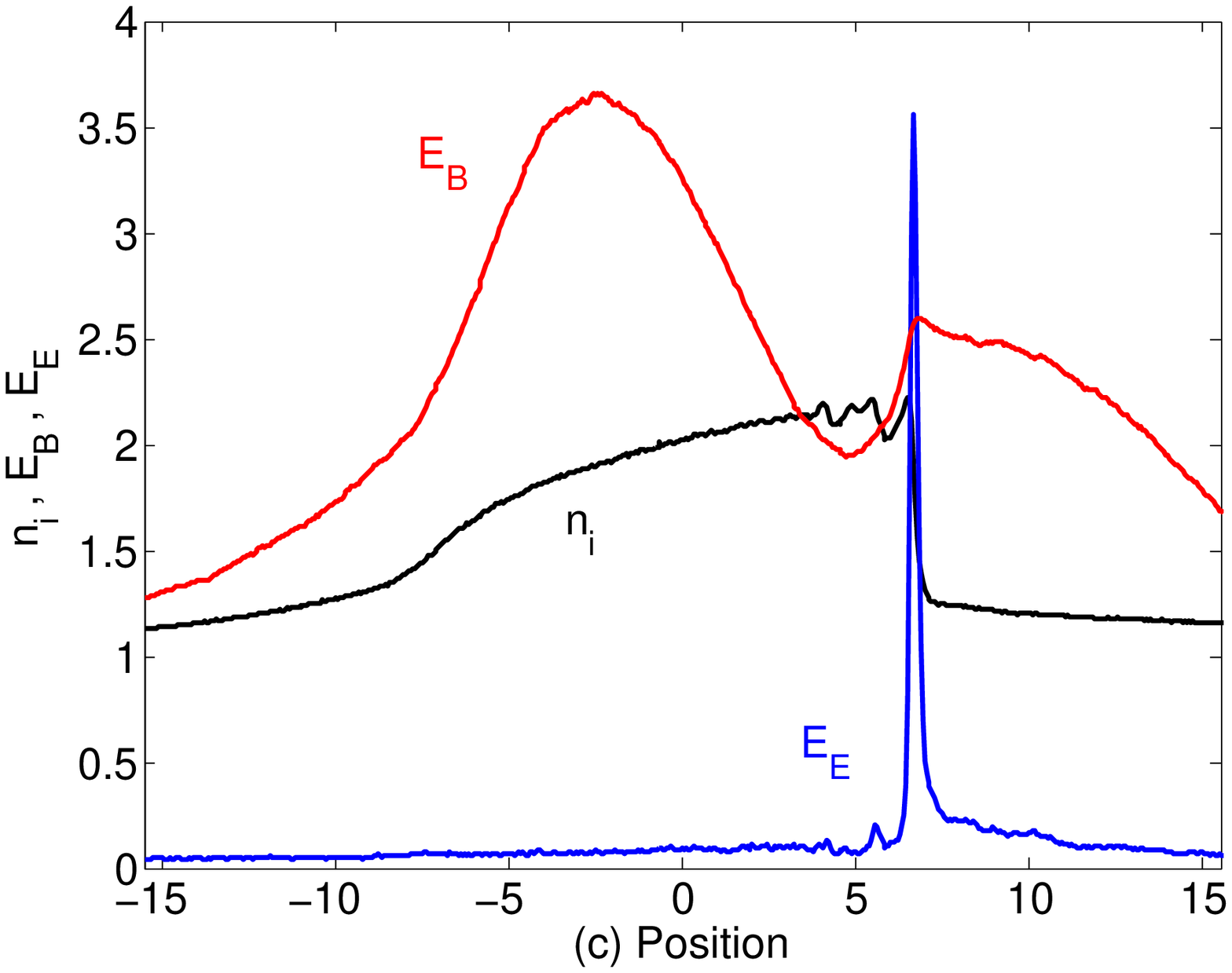}
\caption{Panel (a) shows the y-averaged ion phase space density distribution $f_i (x,v_x)$ and panel (b) the y-averaged electron phase space density distribution $f_e(x,v_x)$. The color scale is linear. Panel (c) shows $n_i (x,t)$ (black curve), the y-averaged magnetic field energy density $E_B(x,t)$ (red curve) and the electric field energy density $E_E(x,t)$ (blue curve). The simulation time is $t = 106$.}\label{phasespace60}
\end{center}
\end{figure}

The plasma phase space density distributions in Fig. \ref{phasespace60}(a,b) demonstrate that the quasi-planar electric field structure in Fig. \ref{electric60}(a) coincides with the location of the hybrid structure. The gradual change of the electron's thermal pressure in the interval $-10<x<0$ results in a weaker ambipolar electric field, which can not be detected in the noise field in Fig. \ref{electric60}. However, the potential difference associated with this electrostatic field, which is spread out over a much larger spatial interval, is sufficient to sustain a change of the ion mean speed between $-10 < x < -5$ that is comparable to that at the hybrid structure. The change in the mean speed expressed in units of the ion acoustic speed is nevertheless smaller in the interval $x<0$. The absence of any steepening of the ion acoustic wave in the left interval indicates that the speed change is less than the local ion acoustic speed. The counterstreaming ion beams in the interval $-15 < x < -8$ do not yield the oblique wave modes, which we find in the interval $7 < x < 15$ in Fig. \ref{electric60}. We attribute this to the Landau damping caused by the much hotter ions of the right-moving ion beam, which delays or suppresses the growth of these waves.

The electric field energy density $E_E(x,t)$ in Fig. \ref{phasespace60}(c) shows a good correlation with the ion density distribution $n_i (x,t)$. The strong peak of $E_B(x,t)$ coincides with the steepest ion density gradient at $x \approx 6.5$ and $E_B (x,t)$ is elevated in the interval $-8 < x < 5$ with the weak density gradient. The gradient of $E_B (x,t)$ is much smaller than in Fig. \ref{phasespace53}(b) and the electron drift speed is thus lower. It is unlikely that the drift instability can be sustained at this time. The ion density behind the hybrid structure is $2n_0$. The magnetic energy density $E_B (x,t)$ is depleted just behind the hybrid structure and reaches its minimum at $x\approx 5$. It has been boosted just ahead of the shock to about 2.5 times its initial value. The anti-correlation between the thermal pressure and $E_B (x,t)$ in the interval $x<0$ is less clear. 

\subsection{Time 491: Towards a fluid shock}

\begin{figure}
\begin{center}
\includegraphics[height=5.8cm]{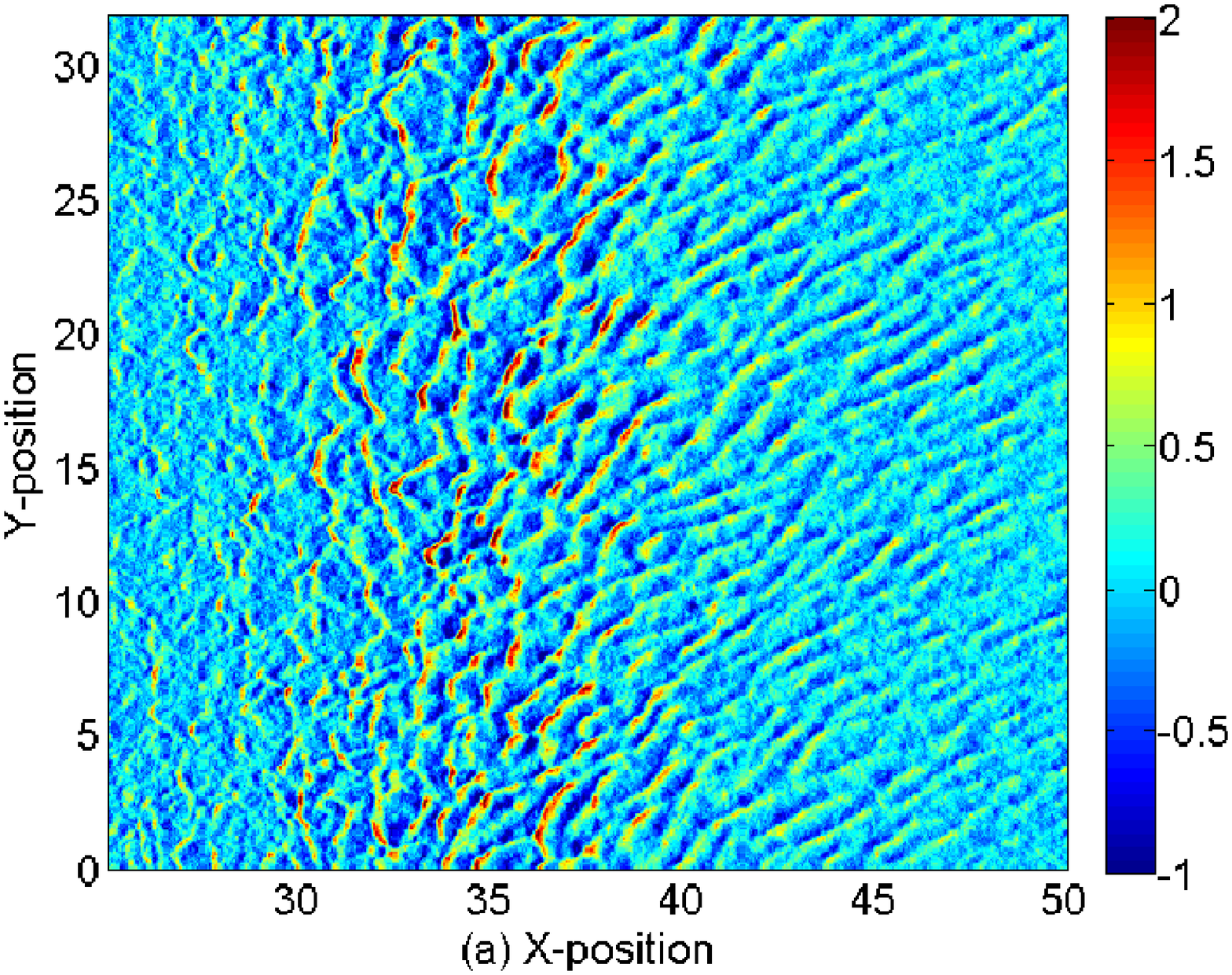}
\includegraphics[height=5.8cm]{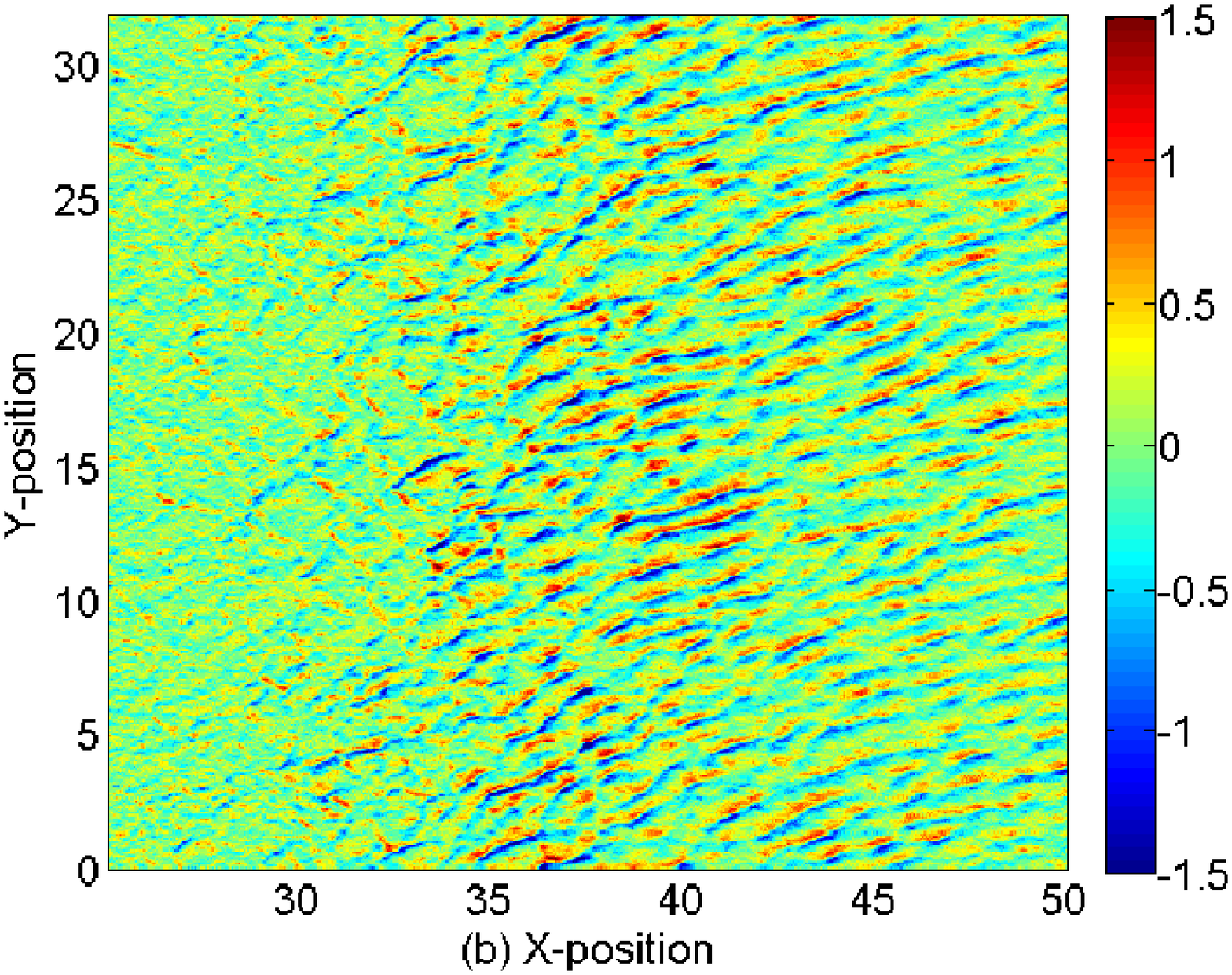}
\caption{The in-plane electric field in units of $10^{-2} m_e c \omega_{pe} / e$: Panel (a) shows $E_x$ and panel (b) shows $E_y$ in a sub-interval of the simulation box. The color scale is linear and $t = 491$.}\label{electric491}
\end{center}
\end{figure}

Figure \ref{electric491} displays the in-plane electric field at $t$=491.
The localized quasi-planar electric field pulse, which sustained the hybrid structure in the interval $x>0$ at earlier times, has been replaced by a broad interval along $x$ with strong wave activity. The waves in the interval $32 < x < 37$ are the strongest ones. Their characteristic wavelength is of the order $0.5\lambda_e$. The characteristic amplitude of the waves decreases with increasing $x>37$ and the angle between their wave vector and the x-axis increases. The oblique waves, which started to develop just ahead of the hybrid structure at $t=106$, have spread out over a spatial interval with a width of $\approx 10 \lambda_e$. 

\begin{figure}
\begin{center}
\includegraphics[height=5.7cm]{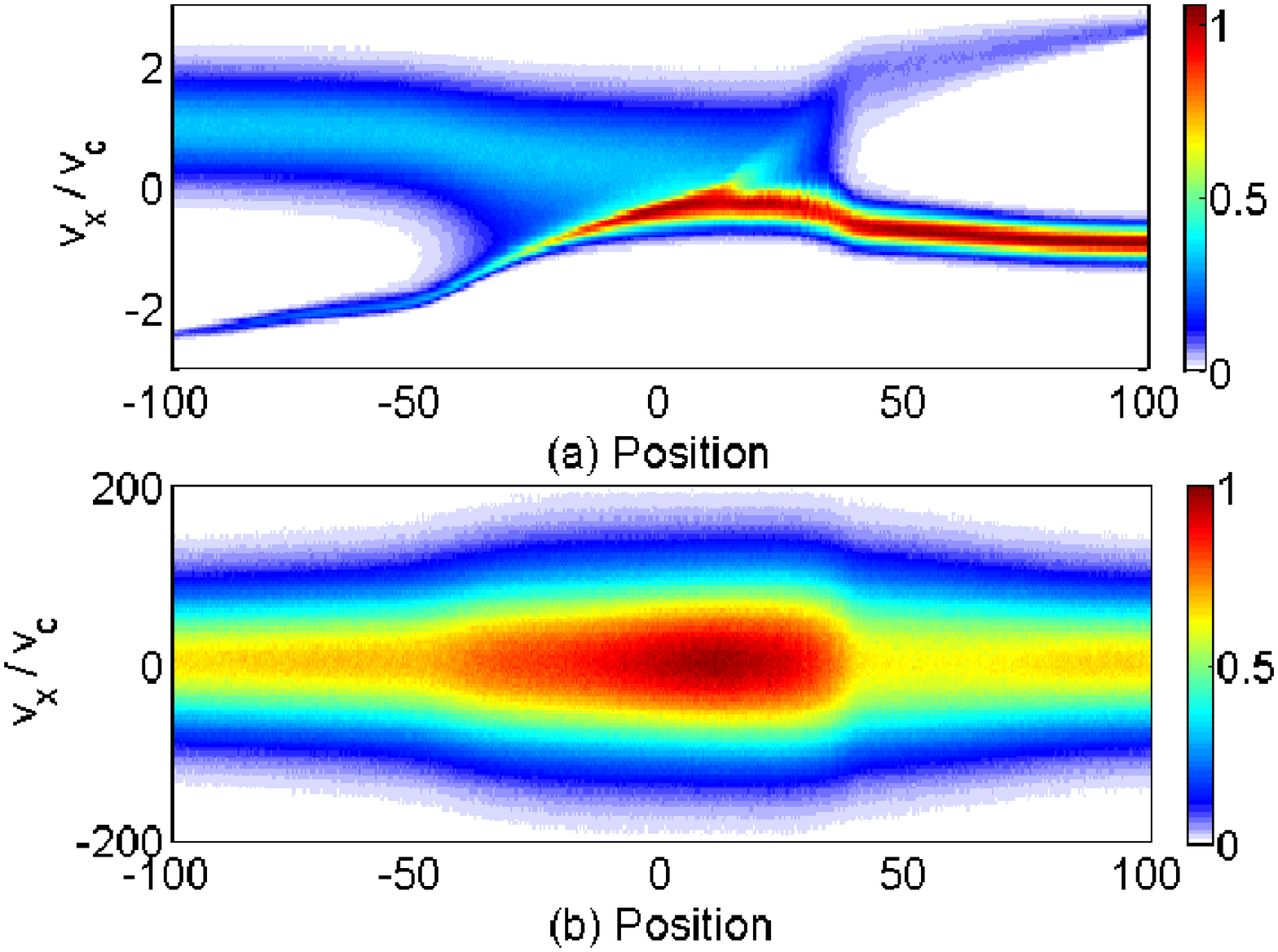}
\includegraphics[height=5.7cm]{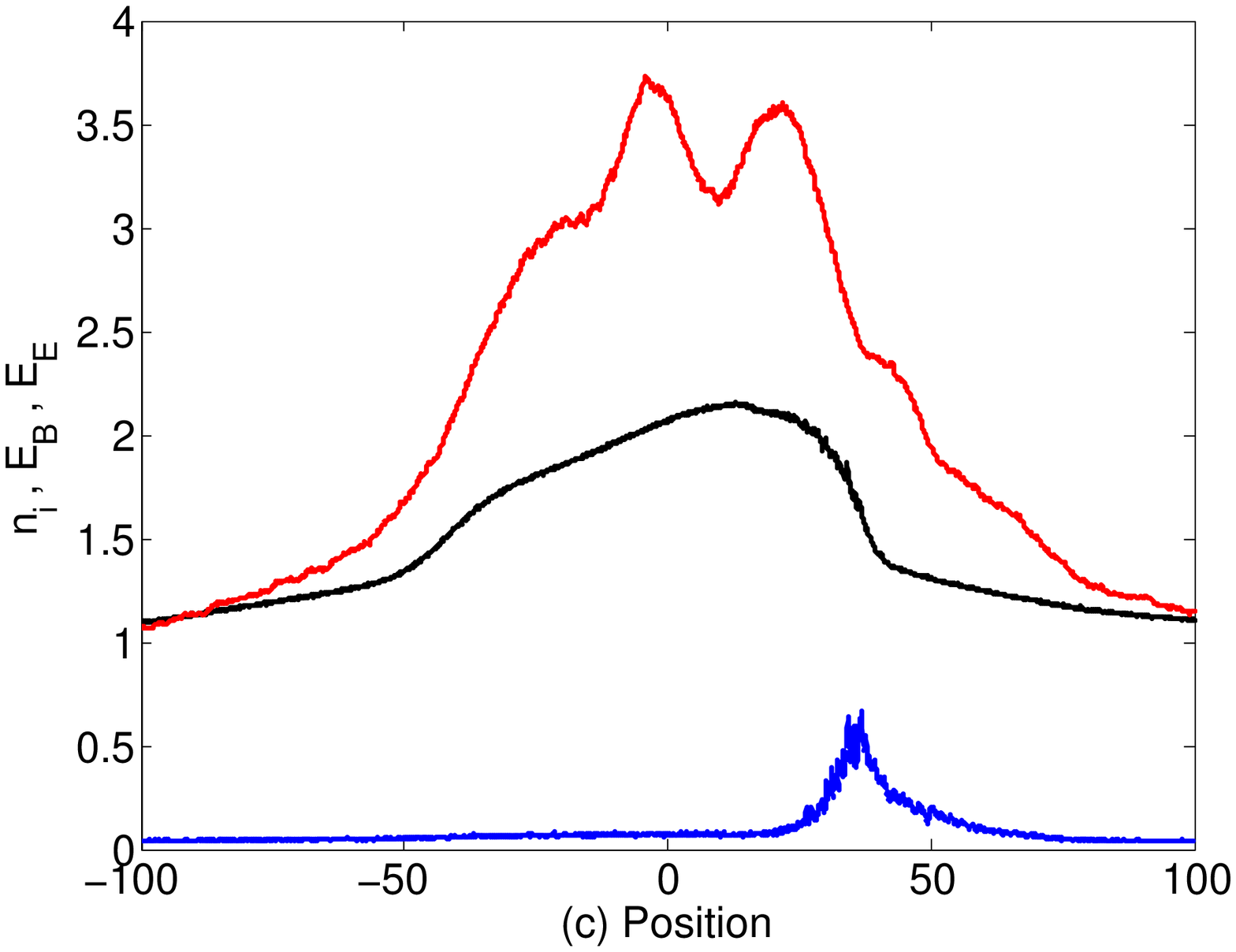}
\caption{Panel (a) shows the y-averaged ion phase space density distribution $f_i (x,v_x)$ and panel (b) the y-averaged electron phase space density distribution $f_e(x,v_x)$. The color scale is linear. Panel (c) shows $n_i (x,t)$ (black curve), the y-averaged magnetic field energy density $E_B(x,t)$ (red curve) and the electric field energy density $E_E(x,t)$ (blue curve). The simulation time is $t = 491$.}\label{phasespace491}
\end{center}
\end{figure}

Figures \ref{phasespace491}(a,b) reveal why the distribution and amplitude of the electrostatic waves in the interval $32 < x < 37$ differs from that of the waves in the interval $x>37$.
Two counterstreaming ion beams are located in the interval $x>37$ in Fig. \ref{phasespace491}(a) and the instability between them drives the oblique waves. The obliquity angle that increases with $x$ reflects the gradual increase of the relative speed between both beams \cite{ForslundC}. Their wave vector would become orthogonal to the shock normal for even higher beam speeds \cite{Kato10}. The strong electrostatic waves within $32 < x < 37$ in Fig. \ref{electric491} are located in the interval, in which the left-moving ions are slowed down before they enter the downstream region at $x\approx 30$. This slowdown takes place over an interval with a width of $\approx 10 \lambda_e$, which is 40 times the width of the hybrid structure in Figs. \ref{electric10.6} and \ref{electric30}. 

Panel (a) of the supplementary movies 2 and 3 animate the phase space density distributions of ions and electrons, respectively. The density $n_i (x)$ of the ions in units of their initial density $n_0$ is shown in the panel (b) of movie 2. Panel (b) of movie 3 animates the electron's thermal energy normalized to the initial one. Both movies cover the interval $0 < t < 491$. They demonstrate that the change from a hybrid structure, which is mediated by a narrow electric field pulse, to the shock with a wide transition layer is gradual. The gradients of the ion density and of the electron's thermal energy are eroded in time. However, the differences between the ion densities and electron temperatures upstream and downstream remain unchanged.

The widening of the interval, across which the left-moving ions are decelerated in Fig. \ref{phasespace491}(a), gives rise to a decreasing magnitude of the ion density gradient. The ion density $n_i (x,t)$ changes from a downstream value 2 to $n_i (x,t)\approx 1.2$ in the overlap layer over an interval with a width $\approx 10 \lambda_e$. The density gradient's modulus for $x<0$ is even lower. The magnetic field energy density reaches now a maximum of $E_B (x,t) \approx 3.5$ at $x\approx 0$ and it decreases monotonically for increasing $|x|>30$. It converges to its initial value at $|x| \approx 100$. The perpendicular magnetic field component is now being compressed in the downstream region, as expected from an MHD shock. The oscillations of $E_B (x \approx 0,t=491)$ indicate though that a steady state has not yet been reached.

The waves, which are driven by the counterstreaming ion beams in the overlap layer, move slowly in the reference frame of the simulation box. They would be growing aperiodically if both ion beams were equally dense \cite{ForslundC,Dieckmann13b}. The downstream region, which expands at the speed $\approx 1.3 \times 10^5$ m$\textrm{s}^{-1}$ in the simulation frame, catches up with them. Figure \ref{Average}(c) shows that the electrostatic waves accumulate ahead of the line that is fitted to the initial expansion speed of the hybrid structure. This accumulation can also be seen from the supplementary movie 4, which animates in time the total ion density during $455 < t < 491$. The oblique density modulations show a lateral motion and the expanding downstream region can catch up with them. The ion density modulations in the downstream region are practically stationary in the shock frame.

The ambipolar electric field is still present at $t=491$, because the thermal pressure gradient of the electrons persists at this time in Fig. \ref{phasespace491}(b) and in movie 3. We find a positive mean electric field within $30 < x < 40$ and a relatively strong $E_B (x,t=491) \approx 2.5$. The electrostatic waves in the interval $32 < x < 37$ might be boosted by the free energy contained in a $\mathbf{E} \times \mathbf{B}$-drift current.  However, the strong electric field oscillations on spatial scales that are comparable to the electron gyroradius imply that this would not be a simple guiding center drift of the electrons.   

\begin{figure}
\begin{center}
\includegraphics[height=5.8cm]{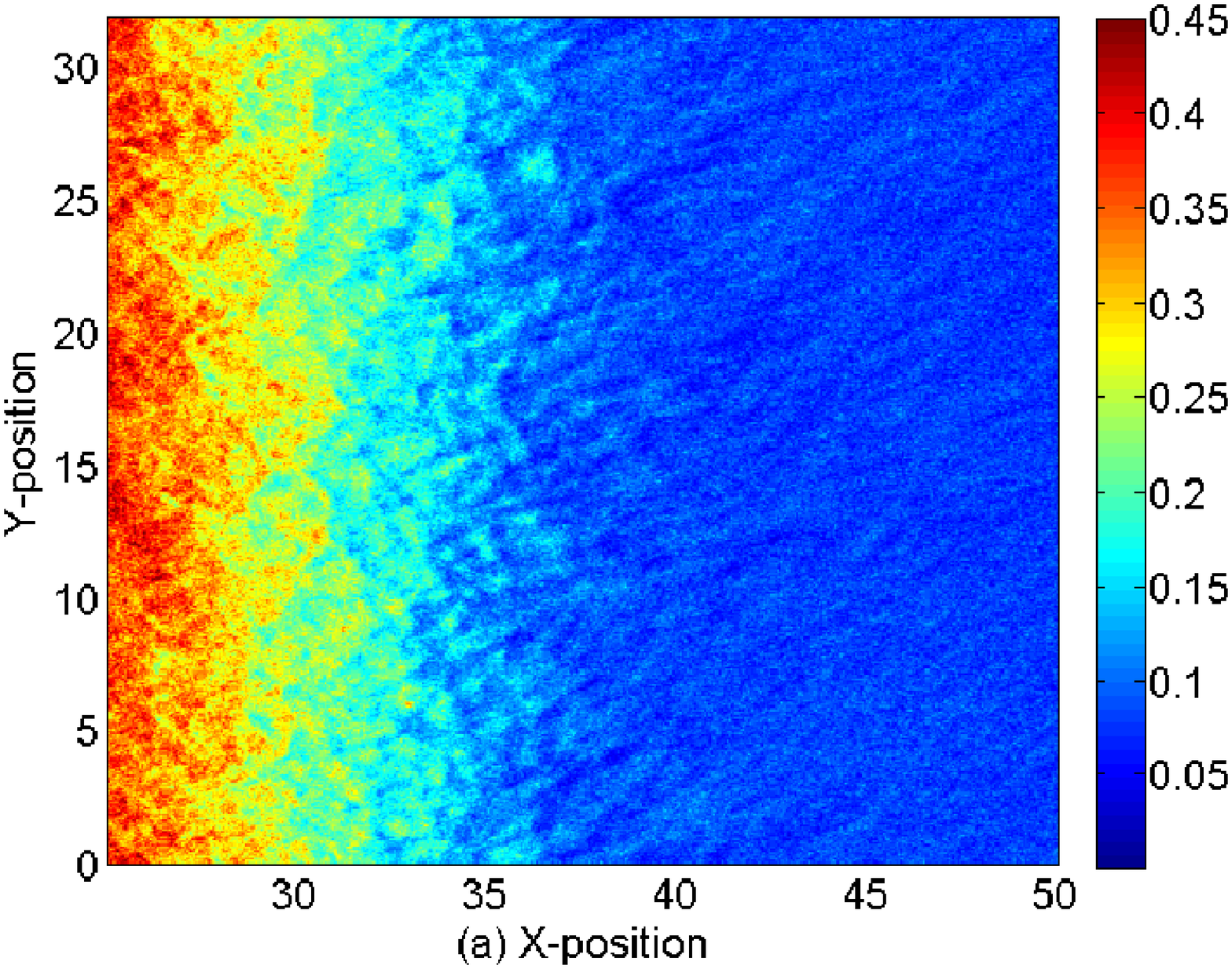}
\includegraphics[height=5.8cm]{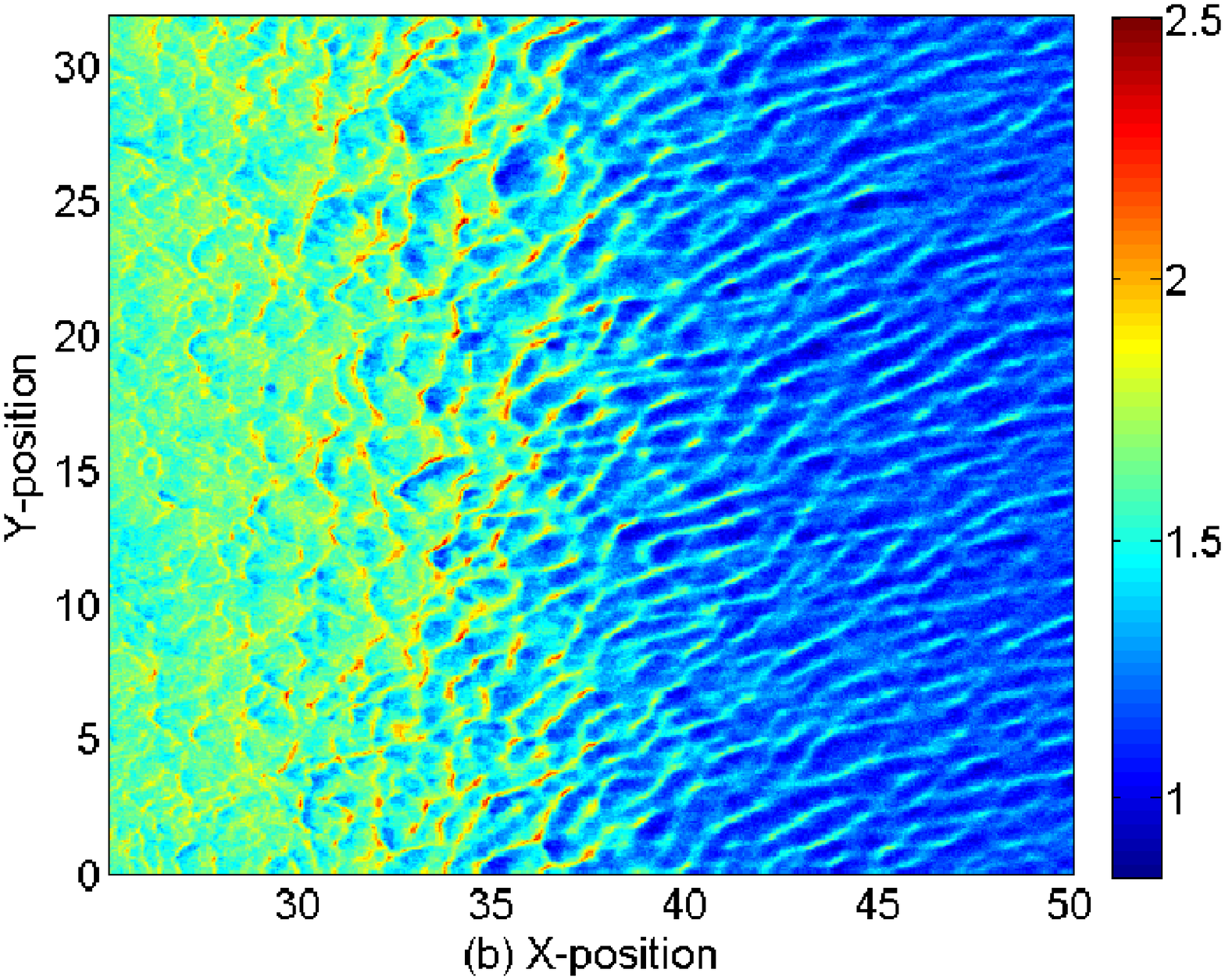}
\caption{Panel (a) shows the ion density of the left beam and panel (b) shows the ion density of the right beam in units of the inital ion number density and in a sub-interval of the simulation box. The color scale is linear and $t = 491$.}\label{IonDensity}
\end{center}
\end{figure}

Figure \ref{IonDensity} shows the spatial density distribution of each ion beam in the interval $25 < x < 50$.  
The left-moving plasma cloud contributes the bulk of the ions, which we can see from its much larger density values. The density of the right-moving ions decreases from a value 0.45 at the left boundary to less than 0.1 at the right one. Its density is spatially almost uniform for $x>40$ and there is thus a continuous outflow of ions from the downstream region into the overlap layer at $x>37$. The density of the left-moving ion beam is about 1 at the right boundary and it increases to a mean value of about 1.6 at the left boundary. This beam reveals density modulations that are correlated with those of the electric field at $t=491$. Figure \ref{IonDensity}(b) shows that the density striations go over smoothly from the overlap layer into those in the interval $32 < x < 37$. The orientation of the striations changes at $x\approx 37$ because they are approximately stationary in the overlap layer and are piled up by the expanding downstream region. The ion density oscillations reach amplitudes that are comparable to the upstream density and they are thus strongly nonlinear. Ion density modulations are visible only in the left-moving beam, which we attribute to its lower temperature. This lower temperature results in a lower thermal pressure and the ion response to the electric field is thus stronger.  

\begin{figure}
\begin{center}
\includegraphics[height=5.8cm]{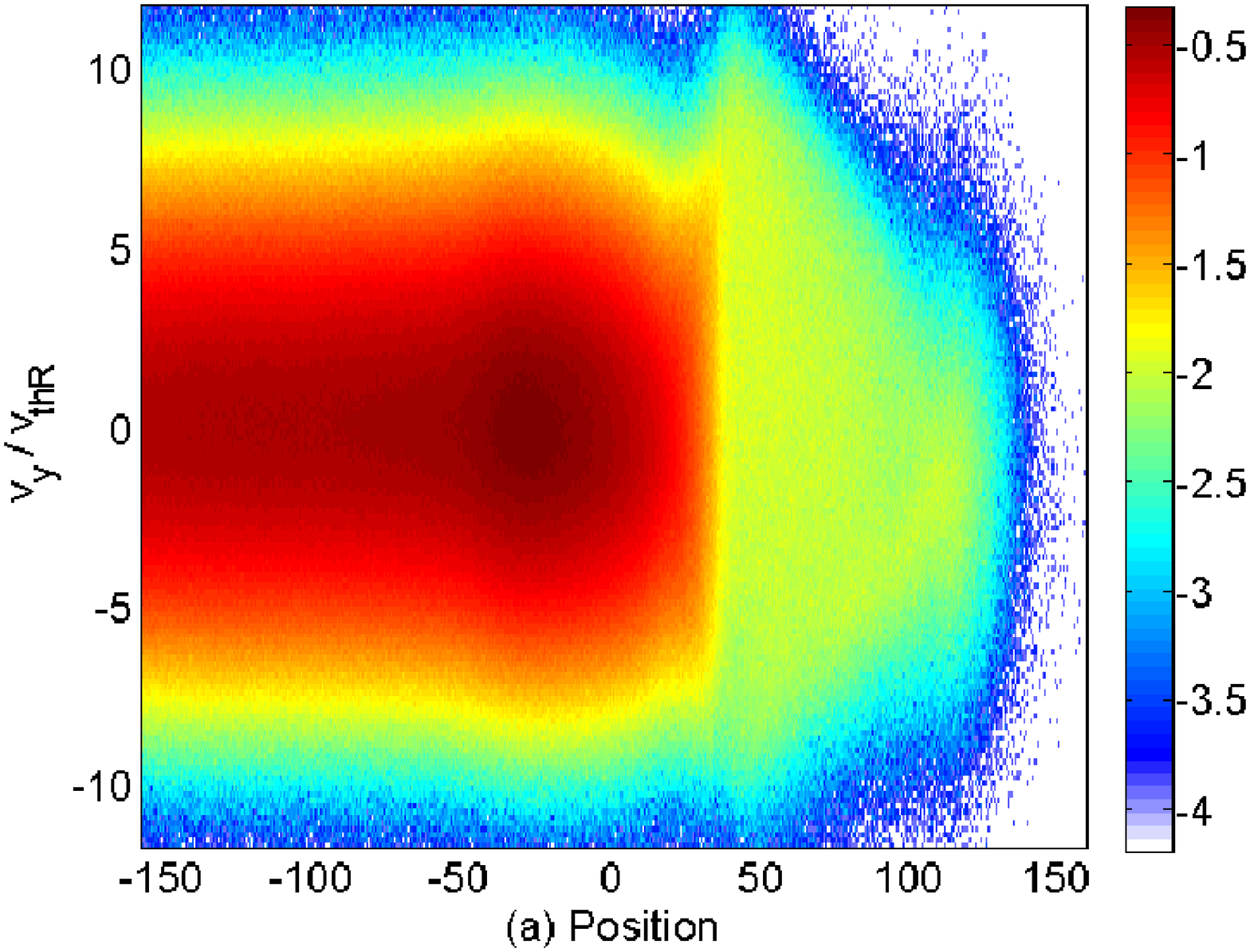}
\includegraphics[height=5.8cm]{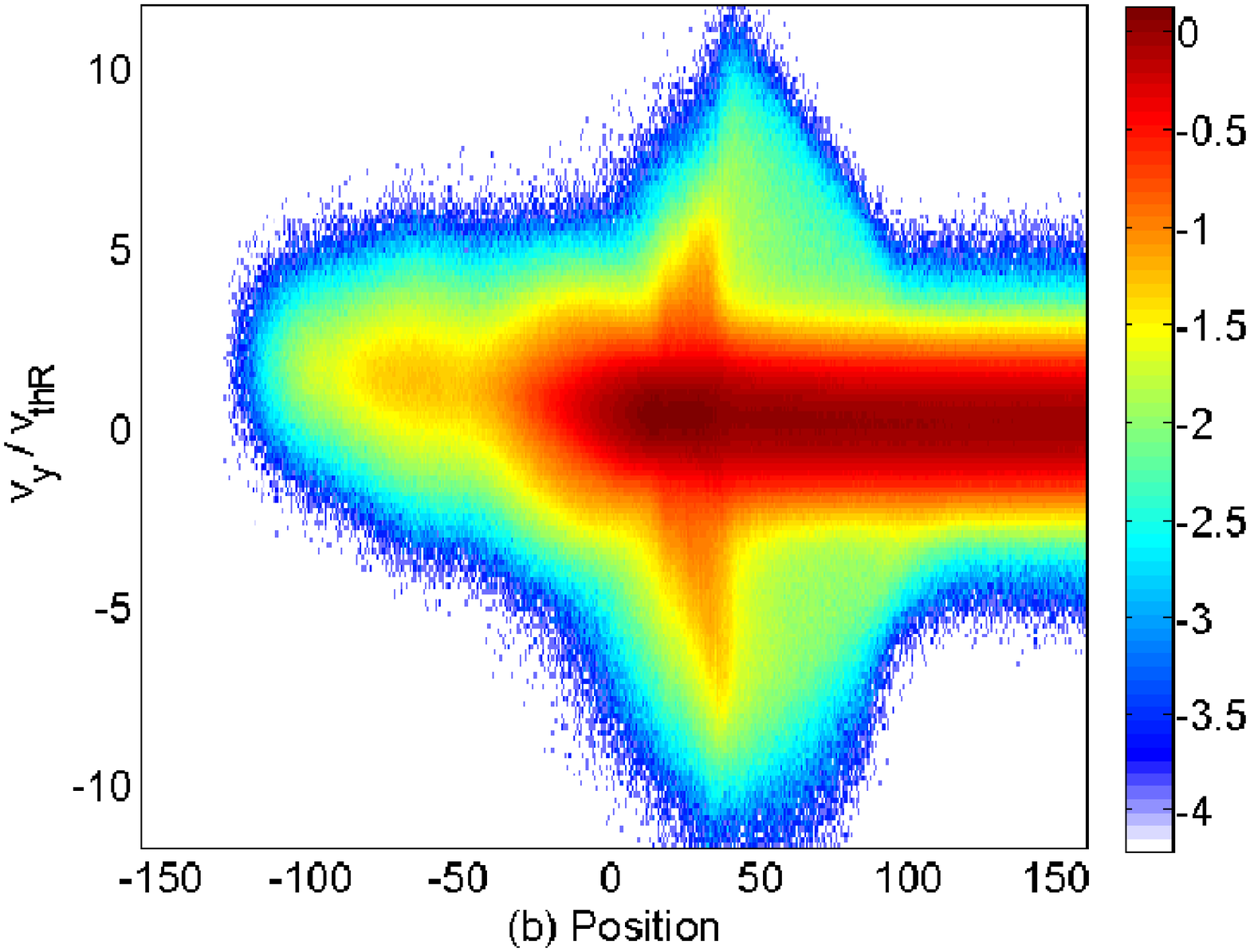}
\caption{Panel (a) shows the y-averaged ion phase space density distribution $f_i (x,v_y)$ of the left beam and panel (b) shows the distribution $f_i (x,v_y)$ of the right beam. The phase space densities are normalized to the peak value in panel (b) and the $v_y$ velocity is normalized to the thermal speed $v_{thR} $ of the right beam. The color scale is 10-logarithmic and $t = 491$.}\label{VYvelocity}
\end{center}
\end{figure}

Figure \ref{IonDensity}(b) reveals strong oblique modulations ahead of the boundary between the downstream region and the overlap layer at $x\approx 30$. Their oblique electric fields (See Fig. \ref{electric491}) should not only modify the ion density, but also the velocity distribution of the ion beams. The effect of the turbulent electrostatic fields on the phase space distribution $f_i (x,v_y,t)$ are revealed by Fig. \ref{VYvelocity}. 
The right-moving ion beam in Fig. \ref{VYvelocity}(a) still shows its initial distribution at $x=-150$. A subtle broadening of the velocity distribution occurs for $-50 < x < 0$. Its phase space density reaches its peak value at $x\approx -30$, it decreases gradually with increasing $x>-30$ and sharply at $x\approx 30$. The ions in the interval $x>30$ correspond to those, which have been accelerated by the double layer component of the hybrid structure. The fastest ions have reached the position $x\approx 130$. The ions of the right-moving beam have not experienced a significant acceleration during the simulation time. The phase space density distribution of the cooler left-moving ion beam shows a more complex distribution. The distribution is the initial one in the interval $x>100$. The dense core population of the ions maintains its peak density until $x\approx 0$ and the density decreases for decreasing values of $x$. The ions that have propagated farthest have reached $x\approx -130$. A triangular structure with a density that is two orders of magnitude below the maximum one is observed in the interval $0 < x < 100$, which reaches the peak speed $\approx 11v_{thR}$. The shape and velocity width of the ion distribution in this interval is similar to that of the much hotter right-moving ion beam in Fig. \ref{VYvelocity}(a), which evidences the onset of the thermalization of both ion populations.

The acceleration of a larger number of ions occurs in the interval $30 < x < 50$ and the density reaches here about 10\% of the peak value. This interval coincides with the one with the strong electrostatic waves. This distribution reaches a peak speed $v_y \approx 6 v_{thR}$, which equals $v_c$. The source of this ion distribution are ions from the core of the ion distribution, which are deflected by the strong oblique waves. The fastest ions in the triagonal structure reach a speed $v_y \approx 12 v_{thR}$. This speed is comparable to that of the fastest ions on the tail of the Maxwellian velocity distribution in the simulation frame of reference. The triangular shape arises because an ion deflection by $\pi / 2$ implies that $v_x \approx 0$. These ions can thus not propagate far upstream within a given time. The lower the deflection angle and, thus, $v_y$, the farther  the reflected ions can move upstream. The triangular ion phase space density distribution in Fig. \ref{VYvelocity}(b) can thus be explained by ion deflection by the electrostatic waves. The lack of electrostatic turbulence in the interval $x<0$ implies that no such structure can be observed in this interval. We note that magnetic effects on the ions are negligible, because the low ion cyclotron frequency $\omega_{ci}$ gives us a small magnetic deflection angle $(\omega_{ci} / \omega_{pi}) t \approx 0.08$ rads for $t=491$.

\section{Summary}

We have modeled with a PIC simulation the collision of two plasma clouds, which have consisted of electrons and ions with a charge-to-mass ratio that corresponds to fully ionized atoms with equal numbers of protons and neutrons. We have chosen Deuterium, because it provides the strongest Landau damping of ion acoustic waves for a given temperature. If we observe an ion acoustic instability for Deuterium ions, then we expect that this instability also develops for heavier ions. The amplitude of the perpendicular magnetic field, which we have introduced into the simulation, was such that it led to a ratio between the electron plasma frequency to the electron cyclotron frequency of 100. The collision speed has been set to just under 900 km/s. These initial conditions have been representative for collisions in laboratory plasmas and in the plasmas close to the outer shell of slow SNR blast shells, like that of RCW86 \cite{Helder11}. The electron temperature of 2.7 keV has been higher than that close to the outer shock of RCW86 and comparable to that in a laser-plasma experiment. The ion temperatures have been realistic for a laser-generated plasma.

The introduction of the weak magnetic field had the purpose to see how strongly this magnetic field is amplified and if it gives rise to plasma structures that can be observed in a laser-plasma experiment. The peak magnetic amplitude in the simulation has exceeded the initial one by a factor 2.5. Even this boosted field has not been strong enough to affect the ion dynamics during the simulation time. Its sole effect has been to introduce a grad-B drift of the electrons, which could trigger a drift instability [43-48]. The introduction of such a weak perpendicular magnetic field will probably not have detectable experimental consequences. 

The simulation shows that unipolar planar electrostatic field pulses develop on electron time scales \cite{Ahmed13,Dieckmann13a,Dieckmann13b}. These are hybrid structures, which are a combination of an electrostatic shock and a double layer \cite{Hershkowitz81}. The selection of different ion temperatures and, thus, different ion acoustic speeds in both colliding plasma clouds yields an asymmetric evolution of both plasmas with respect to their initial contact boundary. This asymmetry is a consequence of the different ion temperatures of both clouds. A sharp ion density change, which is indicative of shocks in laboratory experiments, could only be observed in the plasma with the lower ion acoustic speed. This shock triggers the growth of an ion phase space hole, which transforms the unipolar pulse into a tripolar one. Such a structure may have been observed experimentally \cite{Romagnani08}. 

The tripolar electric field pulse is eventually transformed into a broad layer of turbulent electrostatic fields by an instability between the incoming upstream ions and the shock-reflected ones \cite{Forslund72}. The modulus of the ion density gradient between the downstream region and the overlap layer decreases in response to the ion acoustic instability but the turbulence layer preserved the spatial separation between the downstream region and the overlap layer. The shock has thus not collapsed as in Ref. \cite{Kato10}. The turbulence layer involved spatially localized electrostatic waves with a wide range of angles between their wave vector and the shock normal. These waves have been strong enough to deflect the incoming ions in directions other than that of the shock normal. Such a shock is thus capable of thermalizing the ions of the upstream plasma, as they cross the shock and convect downstream. The electrostatic turbulence takes the role of the collisions in a fluid picture. This turbulence has, however, not been sufficiently strong to yield a full thermalization of the downstream ions. Ultimately the shock formation time may thus depend mainly on the properties of the instability that mediates it \cite{Bret13a,Bret13b,Bret13c}. An electrostatic shock in the formulation by \cite{Hershkowitz81}, which has been observed in Ref. \cite{Ahmed13}, merely introduces another transient step in its formation. 

These turbulent wave fields have also been observed at the Earth's bow shock \cite{Walker04}, but we have to point out that the magnetic field is more important at this shock than at shocks in the interstellar medium. 

The charge-to-mass ratio of the ions we have used here is representative for those in laser-plasma experiments. We can thus estimate the time it takes a shock to form in the laboratory based on the results of our numerical simulation. The turbulent shock transition layer has fully developed at $t\omega_{pi} \approx 500$. It has started at this time to equilibrate the ion speeds along both directions resolved by the simulation. The turbulence layer has a width exceeding 10 electron skin depths. A typical value of the density of the ambient plasma, into which the laser-driven blast shell expands, is $n_0 = 10^{15} \mathrm{cm}^{-3}$. We calculate a formation time $t = 500 \omega_{pi}^{-1} \approx 20$ ns and a characteristic width $\Delta_x = 10 c / \omega_{pe} \approx 1$ mm for the fluid shocks.  

\textbf{Acknowledgements:} ME Dieckmann wants to thank Vetenskapsr\aa det for financial support through the grant 2010-4063. The swedish High Performance Computing Center North (HPC2N) has provided the computer time and support.


\section*{References}
\begin{thebibliography}{10}
\bibitem{Ferriere01} K.M. Ferriere 2001 \textit{Rev. Mod. Phys.} \textbf{73} 103.
\bibitem{Bale} S. D. Bale and F. S. Mozer 2007 \textit{Phys. Rev. Lett.} \textbf{98}, 205001.
\bibitem{Mirabel} I. F. Mirabel \emph{et al.}, 1992 \textit{Nature} (London) \textbf{358}, 215.
\bibitem{Gitomer86}  S.J. Gitomer, R.D. Jones, F. Begay, A.W. Ehler, J.F. Kephart and R. Kristal 1986 \textit{Phys. Fluids} \textbf{29} 2679 
\bibitem{Maksimchuk00} A. Maksimchuk, S. Gu, K. Flippo, D. Umstadter and V.Y. Bychenkov 2000 \textit{Phys. Rev. Lett.} \textbf{84} 4108
\bibitem{dHumieres05} E. d'Humieres, E. Lefebvre, L. Gremillet and V Malka 2005 \textit{Phys. Plasmas} \textbf{12} 062704
\bibitem{Macchi13} A. Macchi, M. Borghesi and M. Passoni 2013 \textit{Rev. Mod. Phys.} \textbf{85} 751
\bibitem{Crow75} J.E. Crow, P.L. Auer and J.E. Allen 1975 1975 \textit{J. Plasma Physics} \textbf{14} 65
\bibitem{Schamel87} C. Sack and H. Schamel 1987 \textit{Phys. Rep.} \textbf{156} 311
\bibitem{Mora05} P. Mora 2005 \textit{Phys. Rev. E} \textbf{72} 056401
\bibitem{Quinn12} K. Quinn \textit{et al} 2012 \textit{Phys. Rev. Lett.} \textbf{108} 135001
\bibitem{Hershkowitz81} N. Hershkowitz 1981 \textit{J. Geophys. Res.} \textbf{86} 3307
\bibitem{Silva04} L.O. Silva, M. Marti, J.R. Davies, R.A. Fonseca, C. Ren, F.S. Tsung and W.B. Mori 2004 \textit{Phys. Rev. Lett.} \textbf{92} 015002
\bibitem{Sarri11} G. Sarri \textit{et al} 2011 \textit{New J. Phys.} \textbf{13} 073023
\bibitem{SarriPRL} G. Sarri 2011 \textit{Phys. Rev. Lett.} \textbf{107} 025003.
\bibitem{Ahmed13} H. Ahmed \textit{et al} 2013 \textit{Phys. Rev. Lett.} \textbf{110} 205001
\bibitem{Remington99} B.A. Remington, D. Arnett, R.P. Drake and H. Takabe 1999 \textit{Science} \textbf{284} 1488
\bibitem{Chen07} M. Chen, Z.M. Sheng, Q.L. Dong, M.Q. He, Y.T. Li, M.A. Bari and J. Zhang 2007 \textit{Phys. Plasmas} \textbf{14} 053102
\bibitem{Romagnani08} L. Romagnani \textit{et al} 2008 \textit{Phys. Rev. Lett.} \textbf{101} 025004
\bibitem{Nilson09} P.M. Nilson \textit{et al} 2009 \textit{Phys. Rev. Lett.} \textbf{103} 255001


\bibitem{Morita10} T. Morita \textit{et al} 2010 \textit{Phys. Plasmas} \textbf{17} 122702
\bibitem{Kuramitsu11} Y. Kuramitsu \textit{et al} 2011 \textit{Phys. Rev. Lett.} \textbf{106} 175002
\bibitem{Kugland12} N.L. Kugland \textit{et al} 2012 \textit{Nat. Phys.} \textbf{8} 809
\bibitem{Habersberger12} D. Habersberger, S. Tochitsky, F. Fiuza, C. Gong, R.A. Fonseca, L.O. Silva, W.B. Mori and C. Joshi 2012 \textit{Nat. Phys.} \textbf{8} 95
\bibitem{Fox13} W. Fox, G. Fiksel, A. Bhattacharjee, P.Y. Chang, K. Germaschewski, S.X. Hu and P.M. Nilson 2013 \textit{Phys. Rev. Lett.} \textbf{111} 225002
\bibitem{Huntington14} C.M. Huntington \textit{et al} 2013 arxiv:1310.3337
\bibitem{Woosley86} S.E. Woosley and T.A. Weaver 1986 \textit{Ann. Rev. Astron. Astrophys.} \textbf{24} 205
 
\bibitem{ForslundA} D. W. Forslund, and C. R. Shonk, Phys. Rev. Lett. 
{\bf 25} 1699 (1970).
\bibitem{ForslundB} D. W. Forslund, and J. P. Freidberg, Phys. Rev. Lett.
{\bf 27} 1189 (1971).
\bibitem{Dieckmann13a} M.E. Dieckmann, H. Ahmed, G. Sarri, D. Doria, I. Kourakis, L. Romagnani, M. Pohl and M. Borghesi 2013 \textit{Phys. Plasmas} \textbf{20} 042111
\bibitem{Jackson60} E.A. Jackson 1960 \textit{Phys. Fluids} \textbf{3} 786
\bibitem{ForslundC} D.W. Forslund, and C.R. Shonk, Phys. Rev. Lett. 
{\bf 25} 281 (1970).
\bibitem{Karimabadi91} H. Karimabadi, N. Omidi and K.B. Quest 1991 \textit{Geophys. Res. Lett.} \textbf{18} 1813
\bibitem{Kato10} T.N. Kato and H. Takabe 2010 \textit{Phys. Plasmas} \textbf{17} 032114
\bibitem{Dieckmann13b} M.E. Dieckmann, G. Sarri, D. Doria, M. Pohl and M. Borghesi 2013 \textit{Phys. Plasmas} \textbf{20} 102112 

\bibitem{Koyama95} K. Koyama, R. Petre, E.V. Gotthelf, U. Hwang, M. Matsuura, M. Ozaki and S.S. Holt 1995 \textit{Nature} \textbf{378} 255
\bibitem{Helder09} E.A. Helder \textit{et al} 2009 \textit{Science} \textbf{325} 719
\bibitem{Raymond09} J.C. Raymond 2009 \textit{Science} \textbf{325} 683
\bibitem{Stockem14} A. Stockem, F. Fiuza, A. Bret, R.A. Fonseca and L.O. Silva 2014 \textit{Sci. Rep.} \textbf{4} 3934

\bibitem{Koehler68} A.M. Koehler 1968 \textit{Science} \textbf{160} 303

\bibitem{Borghesi02} M. Borghesi \textit{et al} 2002 \textit{Phys. Plasmas} \textbf{9} 2214

\bibitem{Sarri10} G. Sarri \textit{et al} 2010 \textit{New J. Phys.} \textbf{12} 045006

\bibitem{Brackbill84} J.U. Brackbill, D.W. Forslund, K.B. Quest and D. Winske 1984 \textit{Phys. Fluids} \textbf{27} 2682

\bibitem{Daughton03} W. Daughton 2003 \textit{Phys. Plasmas} \textbf{10} 3103

\bibitem{Daughton04} W. Daughton, G. Lapenta and P. Ricci 2004 \textit{Phys. Rev. Lett.} \textbf{93} 105004

\bibitem{Forslund72} D. Forslund, R. Morse, C. Nielson and J. Fu 1972 \textit{Phys. Fluids} \textbf{15} 1303

\bibitem{Umeda12} T. Umeda, Y. Kidani, S. Matsukiyo and R. Yamazaki 2012 \textit{Phys. Plasmas} \textbf{19} 042109

\bibitem{Belyaev05} V.S. Belyaev \textit{et al} 2005 \textit{Contrib. Plasma Phys.} \textbf{45} 168

\bibitem{Ghavamian01} P. Ghavamian, J. Raymond, R. C. Smith and P. Hartigan 2001 \textit{Astrophys. J.} \textbf{547} 995

\bibitem{Helder11} E.A. Helder, J. Vink and C.G. Bassa 2011 \textit{Astrophys. J.} \textit{737} 85

\bibitem{Castro13} D. Castro, L.A. Lopez, P.O. Slane, H. Yamaguchi, E. Ramirez-Ruiz and E. Figueroa-Feliciano 2013 \textit{Astrophys. J.} \textbf{779} 49

\bibitem{Berezhko03} E.G. Berezhko, L.T. Ksenofontov and H.J. V\"olk 2003 \textit{Astron. Astrophys.} \textbf{412} L11

\bibitem{Bell04} A.R. Bell 2004 \textit{Mon. Not. Royal Astron. Soc.} \textbf{353} 550

\bibitem{Chapman05} S.C. Chapman, R.E. Lee and R.O. Dendy 2005 \textit{Space Sci. Rev.} \textbf{121} 5

\bibitem{Scholer06} M. Scholer and D. Burgess 2006 \textit{Phys. Plasmas} \textbf{13} 062101

\bibitem{Dawson83} J.M. Dawson 1983 \textit{Rev. Mod. Phys.} \textbf{55} 403

\bibitem{Cook11} J.W.S. Cook, S.C. Chapman, R.O. Dendy and C.S. Brady 2011 \textit{Plasma Phys. Controll. Fusion} \textbf{53} 065006

\bibitem{Brady12} C.S. Brady, A. Lawrence-Douglas and T.D. Arber 2012 \textit{Phys. Plasmas} \textbf{19} 063112

\bibitem{Treumann} R.A. Treumann and W. Baumjohann 1997 \textit{Advanced Space Plasma Physics} (London: Imperial College Press)

\bibitem{Dieckmann06} M.E. Dieckmann, B. Eliasson and P.K. Shukla 2006 \textit{New J. Phys.} \textbf{8} 225 

\bibitem{Niemann13} C. Niemann \textit{et al.} 2012 \textit{Phys. Plasmas} \textbf{20} 012108

\bibitem{Barnes66} A. Barnes 1966 \textit{Phys. Fluids} \textbf{9} 1483

\bibitem{Schamel86} H. Schamel 1986 \textit{Phys. Rep.} \textbf{140} 161

\bibitem{Eliasson06} B. Eliasson and P.K. Shukla 2006 \textit{Phys. Rep.} \textbf{422} 225

\bibitem{Bret13a} A. Bret, A. Stockem, F. Fiuza, C. Ruyer, L. Gremillet, R. Narayan and L.O. Silva 2013 \textit{Phys. Plasmas} \textit{20} 042102

\bibitem{Bret13b} A. Bret, A. Stockem, F. Fiuza, C. Ruyer, L. Gremillet, R. Narayan and L.O. Silva 2013 \textit{J. Plasma Phys.} \textbf{79} 367

\bibitem{Bret13c}  A. Bret, A. Stockem, F. Fiuza, E.P. Alvaro, C. Ruyer, L. Gremillet, R. Narayan and L.O. Silva 2013 \textit{Laser Part. Beams} \textbf{31} 487

\bibitem{Walker04} S.N. Walker, H.St.C.K. Alleyne, M.A. Balikhin, M. Andr\'e and T.S. Horbury 2004 \textit{Ann. Geophys.} \textbf{22} 2291
 
\end{thebibliography}
\end{document}